\newif\ifpnas
\pnasfalse   
\ifpnas
  \documentclass[9pt,twoside,lineno]{pnas-new}
  \templatetype{pnasmathematics} 
\else
  \documentclass[showkeys,aps,prmat,reprint,superscriptaddress,notitlepage,onecolumn]{revtex4-1}
\fi

\ifpnas
\leadauthor{Backofen}

\significancestatement{Very important break through}
\else
\fi

 \usepackage{epsfig}
 \usepackage{latexsym}
 \usepackage{epic}
 \usepackage{eepic}
 \usepackage{psfrag}
 \usepackage{rotating}
 \usepackage{multirow,booktabs}
 \usepackage{color}
 \usepackage{cprotect} 
 \usepackage{miller} 
 \usepackage{xspace} 

\usepackage{graphicx}
 
\usepackage{gensymb} 
 \usepackage{amssymb,amsfonts,amsthm,amsmath}
  \usepackage{stmaryrd} 
  \usepackage{import}

  \usepackage{array}

  \usepackage[normalem]{ulem}

\usepackage{subcaption} 
\renewcommand\vec{\mathbf}
\newcommand{\av}[1]{\ensuremath{\langle #1 \rangle}}
\newcommand{\avT}[1]{\ensuremath{\langle #1 \rangle_T}}
\newcommand{\avC}[1]{\ensuremath{\langle #1 \rangle_{\rm C}}}

\newcommand{\uu}{\ensuremath{\vec{u}}\xspace}

\newcommand{\nueff}{\ensuremath{\nu_{\mathrm{eff}}}\xspace}

\newcommand{\GO}{\ensuremath{\Gamma_0}\xspace}
\newcommand{\GT}{\ensuremath{\Gamma_2}\xspace}
\newcommand{\GF}{\ensuremath{\Gamma_4}\xspace}
\newcommand{\tGO}{\ensuremath{\tilde{\Gamma}_0}\xspace}
\newcommand{\tGT}{\ensuremath{\tilde{\Gamma}_2}\xspace}
\newcommand{\tGF}{\ensuremath{\tilde{\Gamma}_4}\xspace}

\newcommand{\Lap}{\ensuremath{\Delta}\xspace}

\newcommand{\activity}{\ensuremath{\alpha_{\rm act}}\xspace}
\newcommand{\nuact}{\ensuremath{\nu_{\rm act}}\xspace}
\newcommand{\nupas}{\ensuremath{\nu_{\rm pas}}\xspace}

\newcommand{\rmic}{\ensuremath{r_{\rm MIC}}\xspace}

\usepackage{tikz}
\usepackage{grffile}

\begin{document}

\title{Self-organized flows break morphological symmetry in active/passive systems}


\ifpnas
\else
  \author{Rainer Backofen}
  \affiliation{Institute of Scientific Computing, Technische Universit\"at Dresden, 01062 Dresden, Germany}
  \author{Axel Voigt}
  \affiliation{Institute of Scientific Computing, Technische Universit\"at Dresden, 01062 Dresden, Germany}
  \affiliation{Center for Systems Biology Dresden, Pfotenhauerstr. 108, 01307 Dresden, Germany}
  \affiliation{Cluster of Excellence, Physics of Life, Technische Universit\"at Dresden, Arnoldstr. 18, 01307 Dresden, Germany}
\fi

\begin{abstract}
We consider a phase-separating mixture of active and passive fluids and explore morphological asymmetries of the emerging dominantly bicontinous dynamic emulsion. Two-dimensional numerical simulations reveal that the geometric and topological asymmetries can solely be explained by self-organized flows in the active region. As in inertial turbulence an inverse energy cascade in the active region leads to the formation of condensates. The size of these mesocales vortices is determined by the locally available space in the emulsion. As these condensates accumulate energy they impact the fluctuation of the surrounding interface and thus form a tight coupling between the flow field and the dynamic morphology. While explored for active/passive systems the symmetry-breaking mechanism can be generalized to heterogeneous active systems and proposes a way to control the morphology of various functional soft materials. 
\end{abstract}

\maketitle

\section{Introduction}

Interfaces of phase-separating fluid mixtures determine their morphology and thus are key to the creation of diverse functional soft materials throughout biology, physics and materials science. Next to approaches which modify the properties of the interface in order to arrest coarsening and control the morphology, e.g., using chemical reactions \cite{zwicker2015emulsion,nakashima2021active,datt2025fluid}, interfacial jamming \cite{stratford2005colloidal,B807312K,aland2011continuum,aland2012particles} 
or active geometric interfacial terms \cite{wittkowski2014scalar,Nardini2017activeB+,Tjhung2018active,al2023chiral,porrmann2024shape}, also bulk properties of the mixture can be used to tune the interface and thus the morphology. Recent studies on mixtures of active and passive fluids have shown that the active region can shape the interface and lead to dynamic emulsions \cite{Pattesonetal_NC_2018,adkins2022interface,Zhangetal_PNAS_2022,zhao2024asymmetric}. Experiments on mixtures of active microtubule-based liquid crystals and passive isotropic fluids \cite{adkins2022interface,zhao2024asymmetric} demonstrate that activity in the bulk phase can arrest phase-separation and strongly enhance interfacial fluctuations leading to asymmetry in the emerging dynamic emulsions. The interfacial fluctuations lead to preferential passive droplets within the active material, which has been associated with elastic forces emerging from the liquid crystalline degrees of freedom in the active region \cite{zhao2024asymmetric}. As a result, nematic elasticity has been proposed as a way to control this asymmetry and to form metamaterials with tunable microstructures. We here propose that the observed asymmetry can also be explained without explicitly requiring liquid crystalline order. Self-organized flows within the active region are sufficient to establish asymmetry in active/passive fluid mixtures. 

Active systems are prone to experience instabilities and self-organization phenomena, thus developing correlated collective flows that can become spatiotemporally chaotic. This makes them analogous to inertial turbulence at a descriptive level \cite{Alertetal_ARCMP_2022} and a huge effort has been made to identify similarities and discrepancies between inertial and active turbulence in one-phase systems
\cite{Alertetal_ARCMP_2022,Bratanovetal_PNAS_2015,Mukherjeeetal_NP_2023,Backofenetal_PNAS_2024}. At large spatial scales a transition between features of active turbulence without and with scaling characteristics of inertial turbulence have been related to characteristics of hyperuniformity, the anomalous suppression of density fluctuations at large spatial scales \cite{Tor18}. The transition occurs at a cross-over from hyperuniformity to non-hyperuniformity and anti-hyperuniformity of the vorticity autocovariance and depends on the strength of active forcing \cite{Backofenetal_PNAS_2024}. Within (quasi-)two-dimensional systems this transition relates to the accumulation of energy at the largest scale in form of the formation of a condensate. In this condensate phase active and inertial turbulence become reminiscent and the size of the condensate is solely determined by the size of the domain. Experiments confirm this behavior for bacterial suspensions \cite{PhysRevLett.110.268102,perez2025bacteria}.

Extending these investigations from one-phase to two-phase systems leads to new phenomena. Here the inverse energy cascade in two-dimensional inertial turbulence can be blocked. This happens at
the length scale at which inertial and interfacial-tension forces balance, the so-called Kolmogorov-Hinze (KH) scale \cite{Hinze_AICHE_1955}. Larger domains become unstable because they are susceptible to deformation and disintegration due to the fluctuating pressure of the surrounding phase. This leads to arrested coarsening \cite{PhysRevA.23.3224,Bertetal_PRL_2005,perlekar2017two}. We will demonstrate that this is also true for active fluid mixtures. The arrested state forms a dynamic emulsion. Its morphology has a strong impact on the flow field and again condensates are formed. But the size of the condensates are no longer determined by the domain size but the new spatial scale defined by the morphology. We will show that this coupling between the morphology and the accumulation of energy is sufficient to establish interface fluctuations which lead to asymmetries in geometric and topological properties of two-phase systems. 

This broadens the applicability, as nematic elasticity is no longer required to form a dynamic asymmetric emulsion. In cell biology this mechanism might be one ingredient to explain the coupling between active processes within the cell and the self-organization of biomelocular condensates \cite{hyman2014liquid,alberti2025current}. Two-phase suspensions with internal boundaries separating different types of bacteria \cite{ajesh2022bacteria} provide another example. And even heterogeneous mixtures of cells with internal boundaries separating different cell types 
\cite{Gibbsetal_Science_2008,Buddingetal_JB_2009,Pattesonetal_NC_2018} fall within this class of problems. All, at least within certain parameter regimes, can be considered as active phase-separating fluid mixtures with long-ranged hydrodynamic flows. In order to shed light on these phenomena we consider the simplest setting of active two-phase suspensions with internal boundaries, by addressing one active and one passive region and use a momentum-conserving extension of the Navier-Stokes (NS) equations, known as the generalized Navier-Stokes (GNS) equations \cite{Slomkaetal_EPJSP_2015} to model the active region and the classical NS equations for the passive region. Both are coupled by the Cahn-Hilliard (CH) equations to build the two-phase system. We explore this system using two-dimensional simulations within a symmetric active/passive setting under low Reynolds number and only vary the strength of activity.

\begin{figure}[htb!]
  \noindent
  \begin{center}
  \begin{tabular}{cccc}
A.0 \includegraphics*[width=0.2\textwidth]{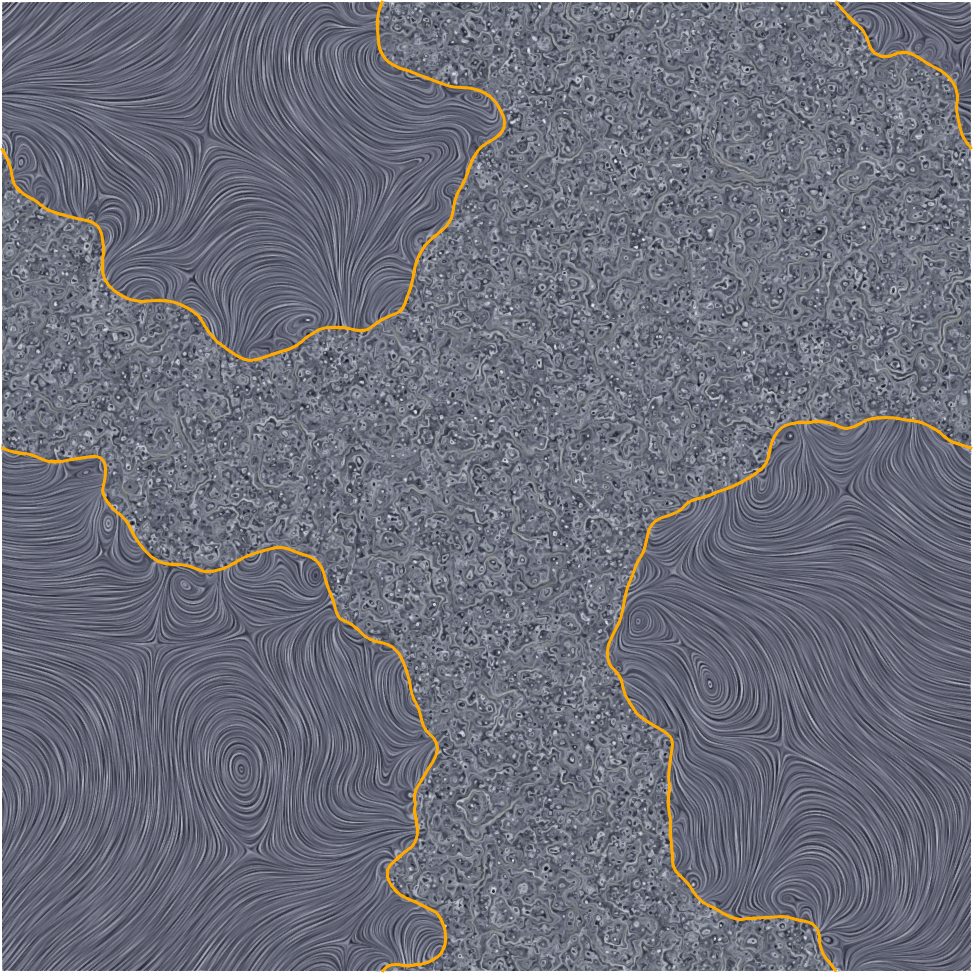}                                                                        
&A.1 \includegraphics*[width=0.2\textwidth]{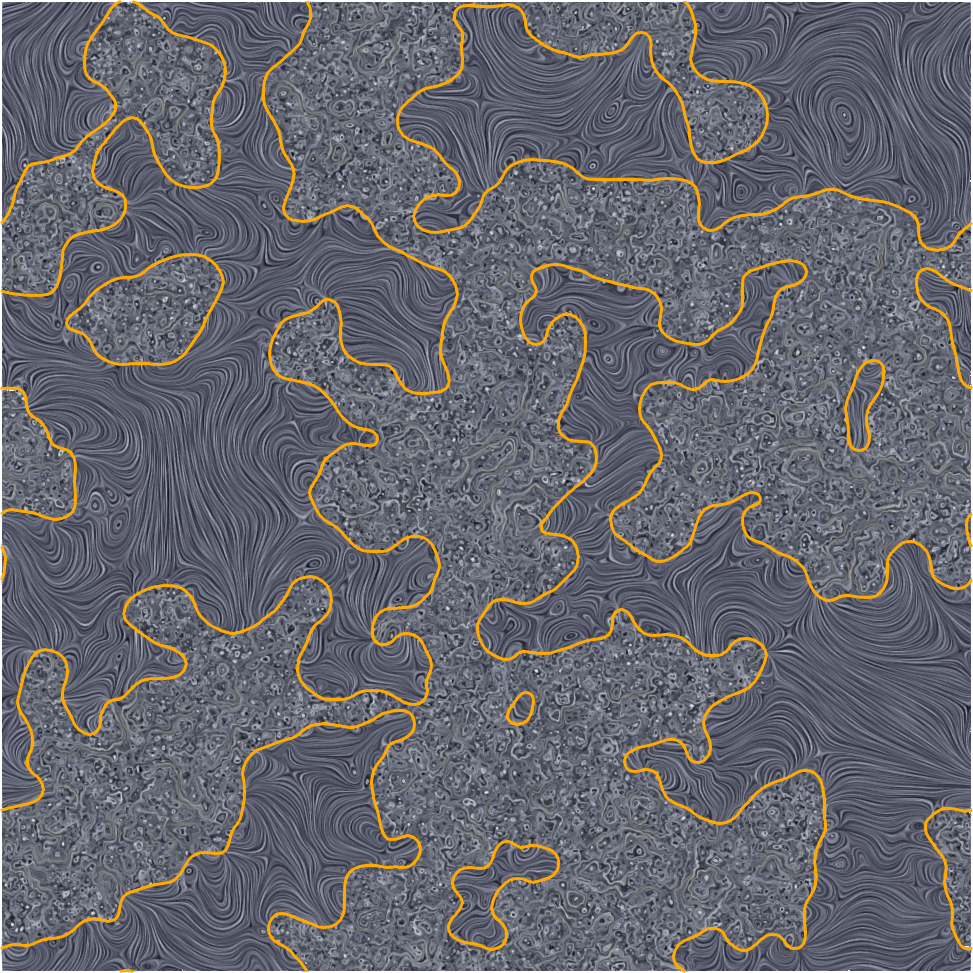}                                                                      
&A.2 \includegraphics*[width=0.2\textwidth]{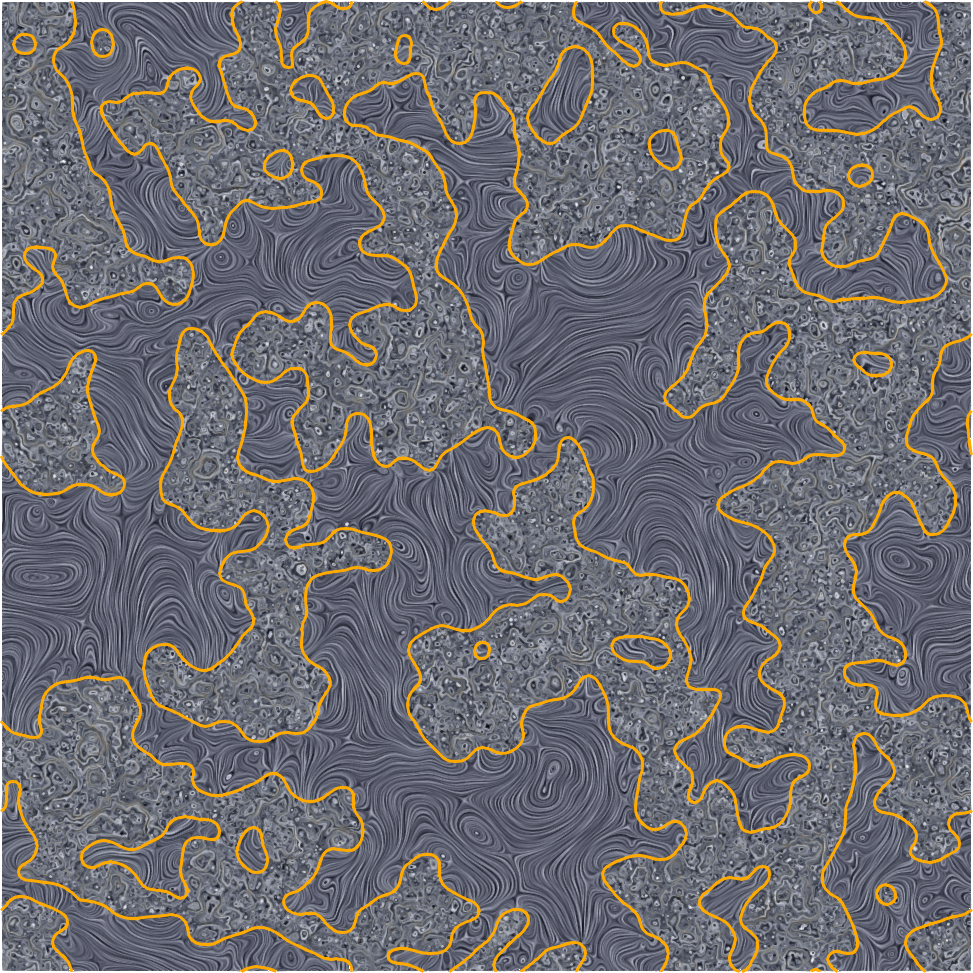}                                                                      
&A.3 \includegraphics*[width=0.2\textwidth]{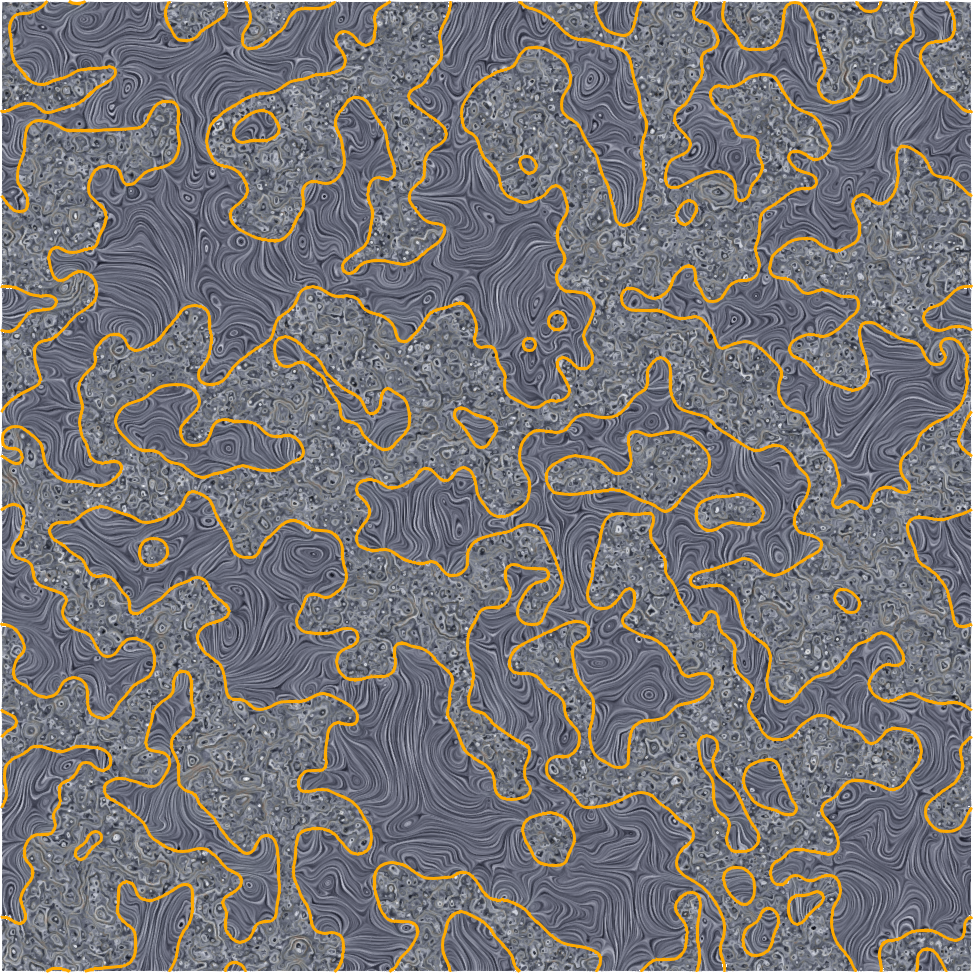}                                                                  
  \end{tabular}
\end{center}
\begin{center}
\begin{tabular}{ccc}
   B \includegraphics*[width=0.3\textwidth]{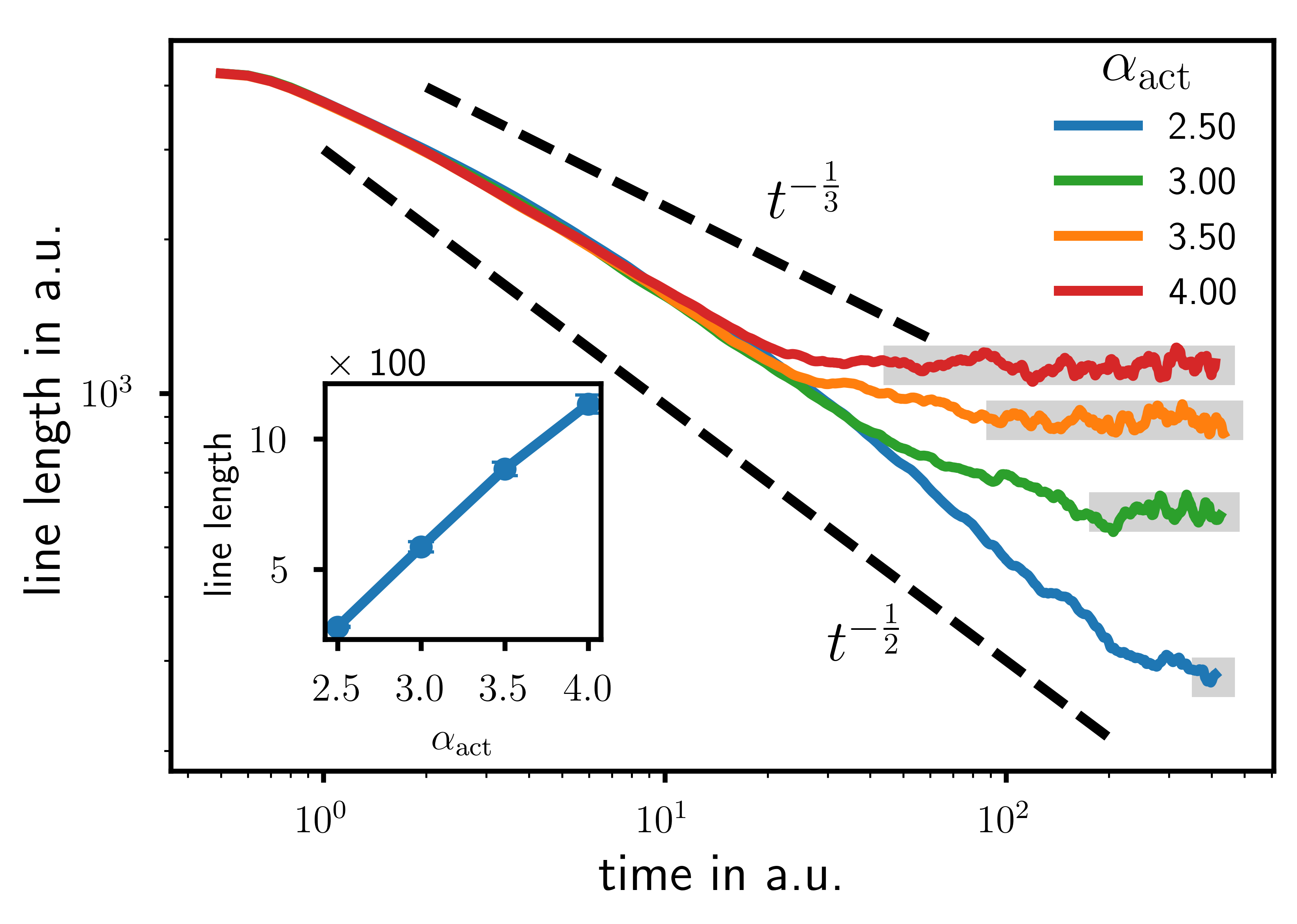} 
    &C \includegraphics*[width=0.3\textwidth]{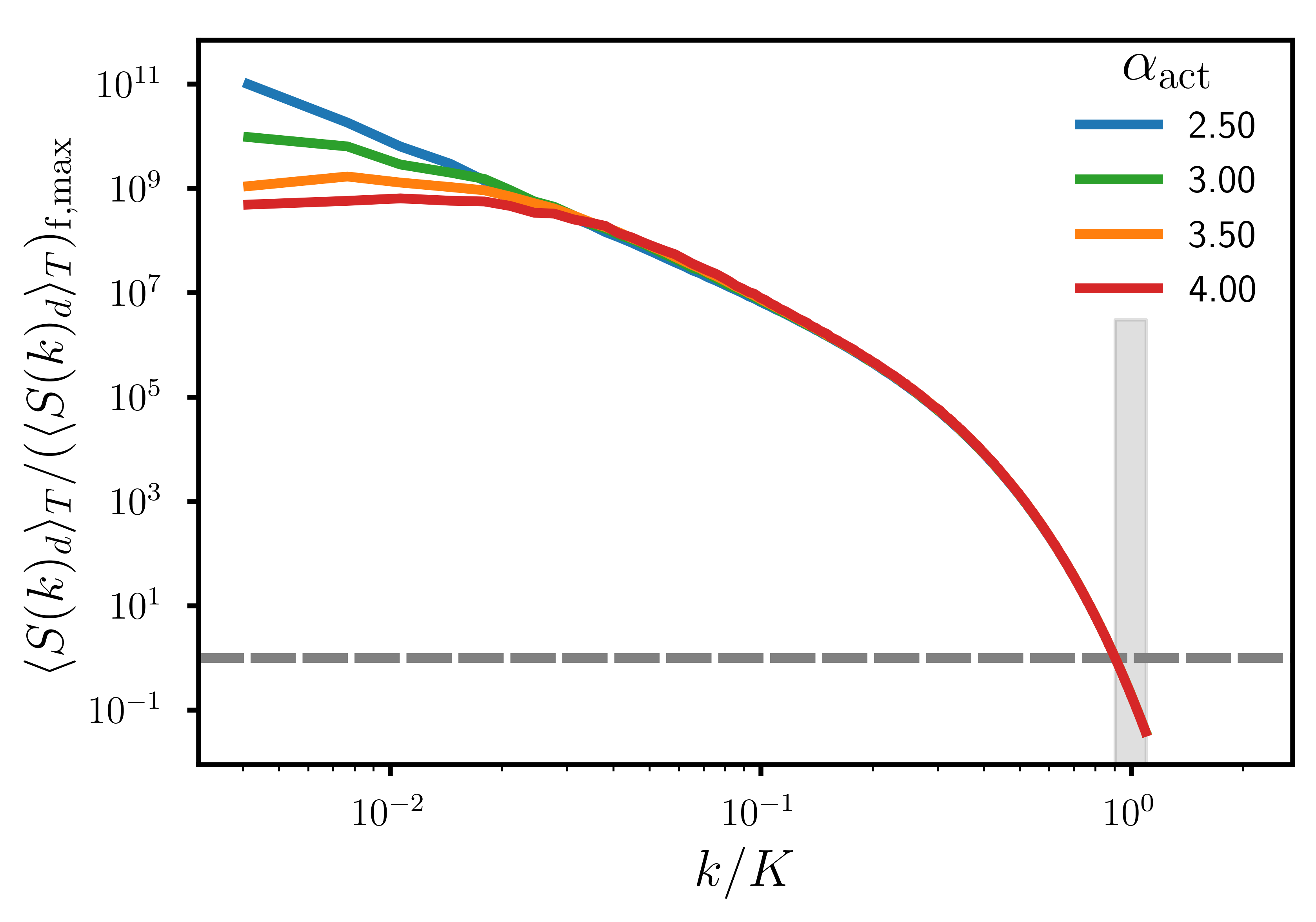} 
  & D
    \includegraphics*[width=0.3\textwidth]{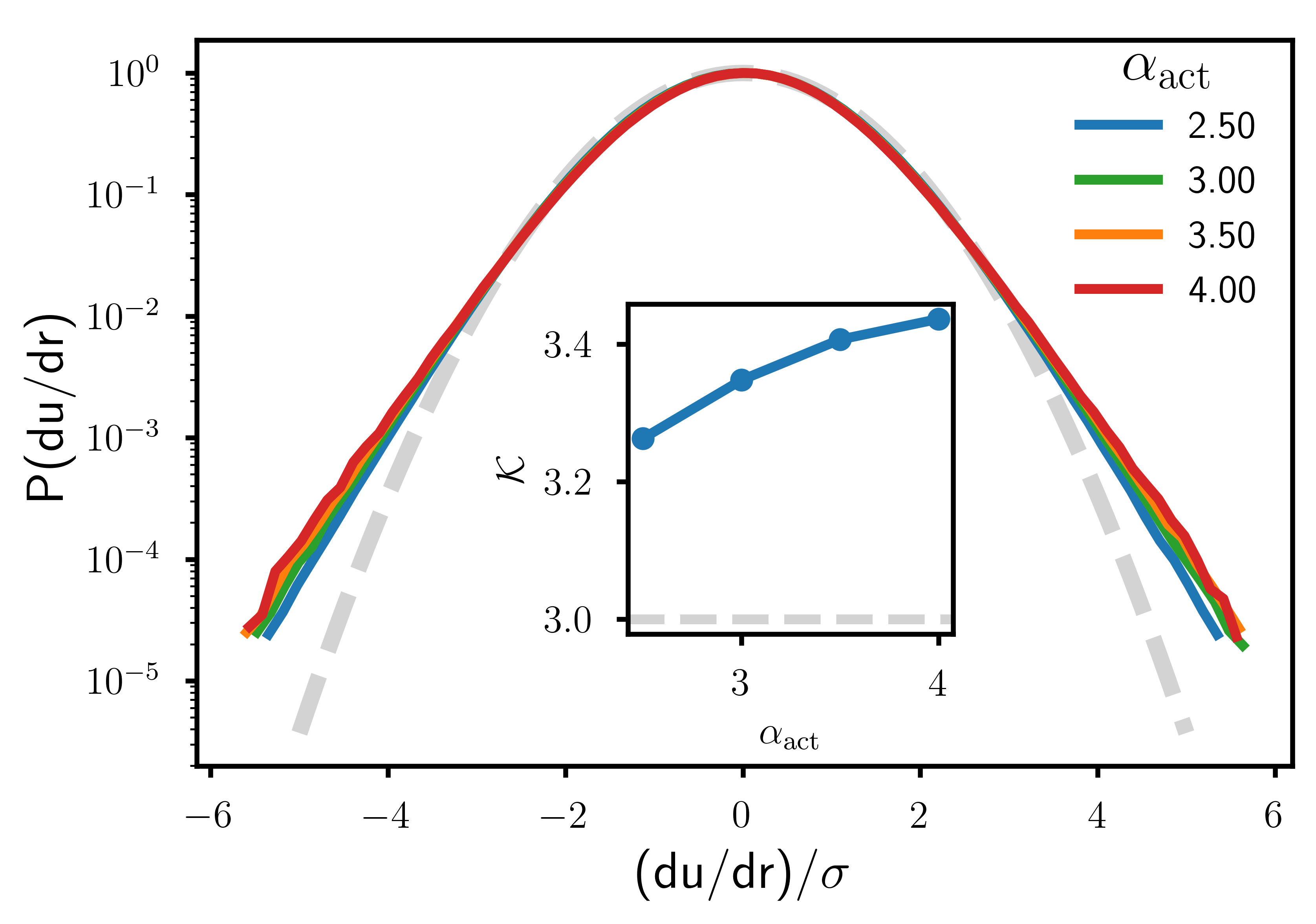} \\
  E \includegraphics*[width=0.3\textwidth]{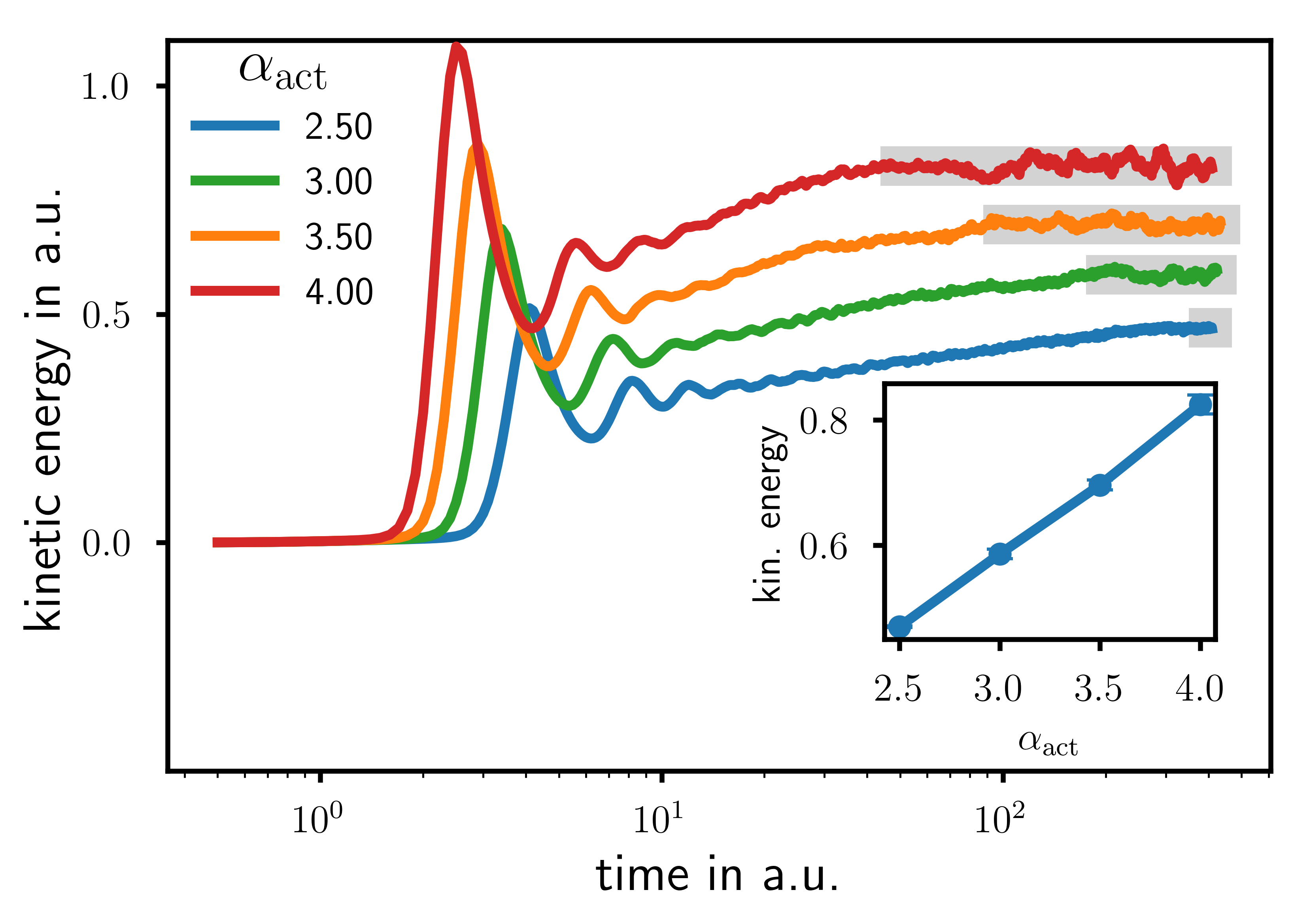}
    & F \includegraphics*[width=0.3\textwidth]{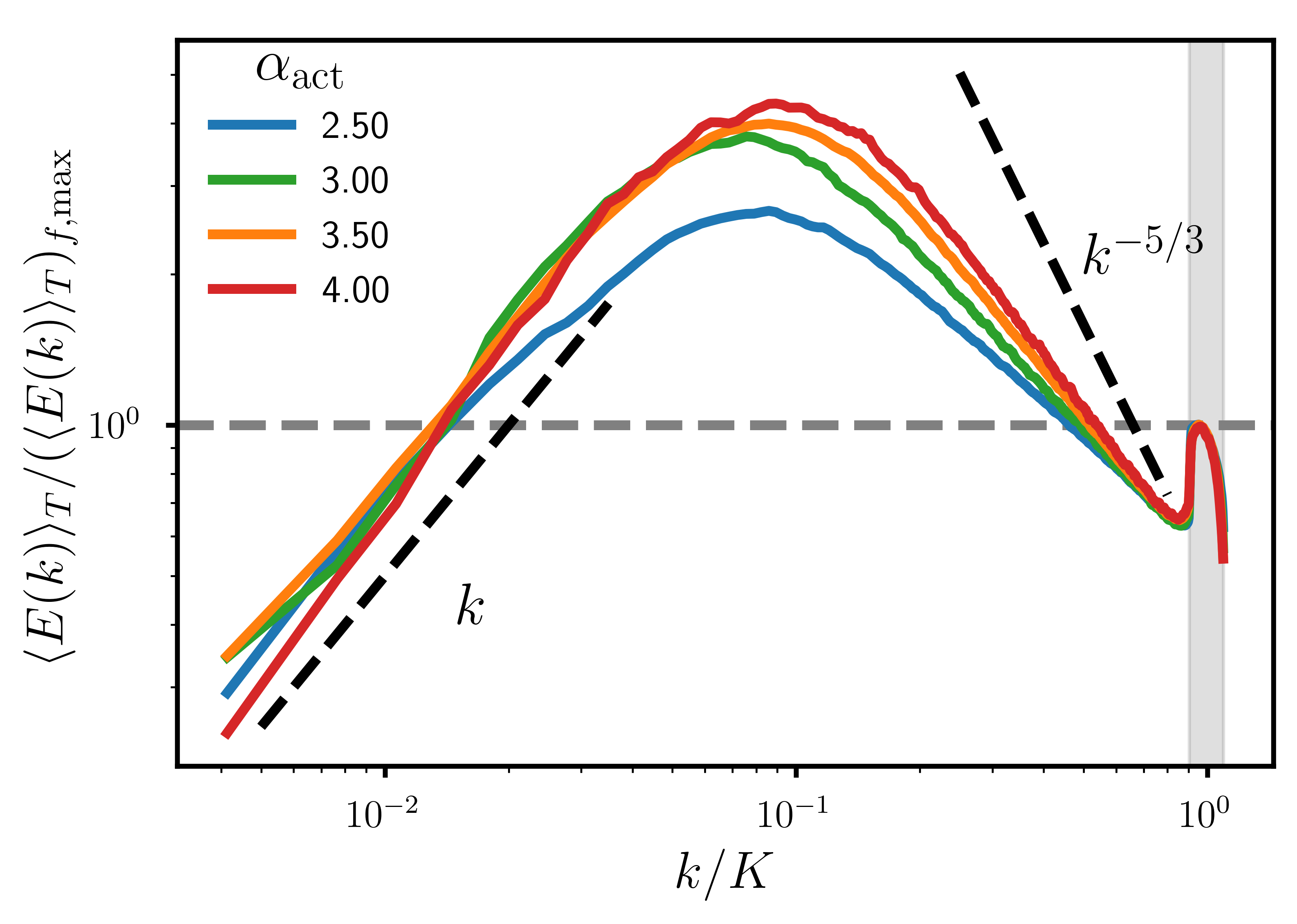}
  &G \includegraphics*[width=0.3\textwidth]{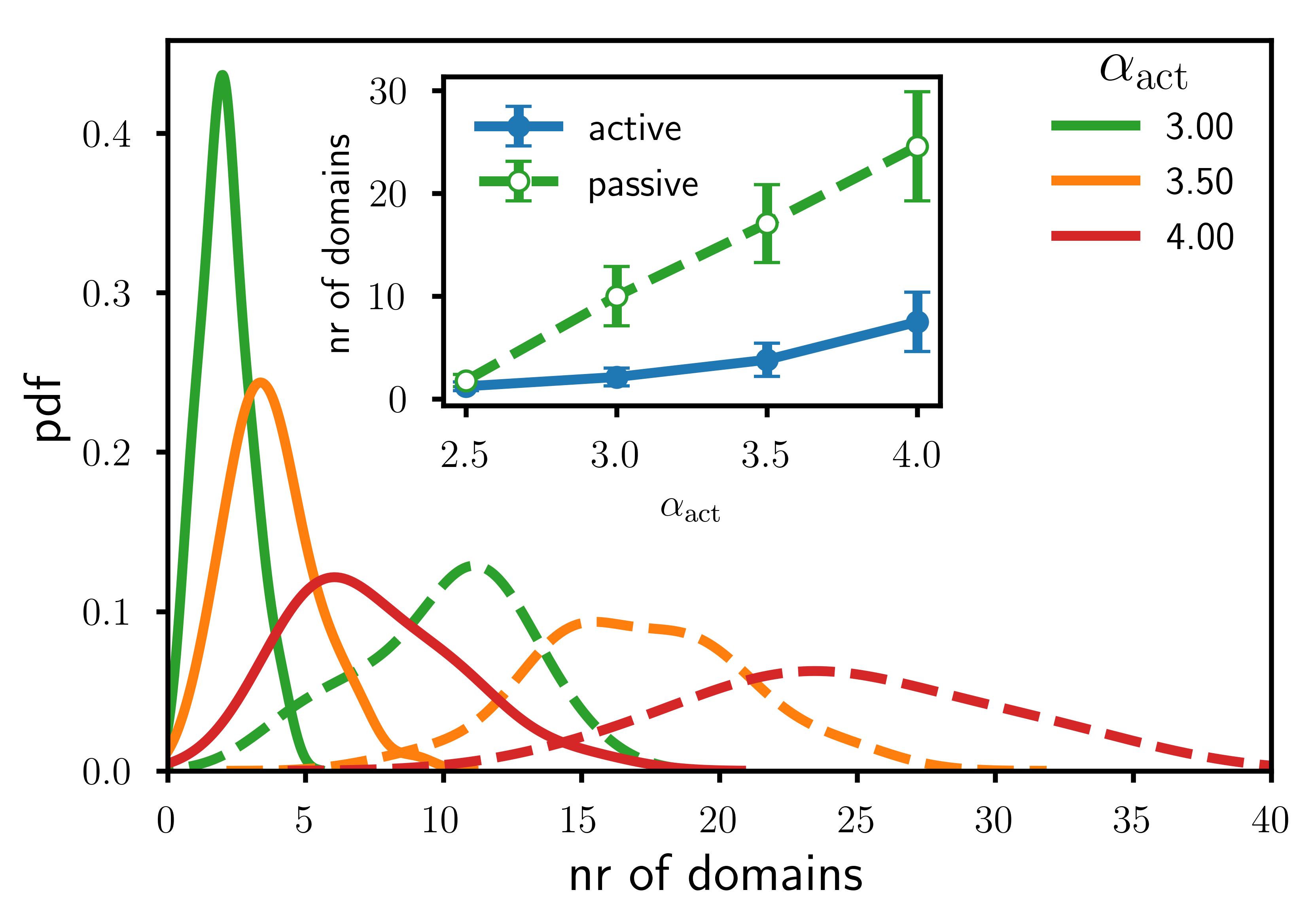}
\end{tabular}
\end{center}
  \begin{center}
    \begin{minipage}{0.95\textwidth}
      \caption[short figure description]{\label{fig:dynEq}
\textbf{A.0-A.3:} LIC (line intensity convolution) visualization of a time instance
of a two-phase system in dynamic equilibrium state at activity $\activity = 2.5, 3, 3.5,$ and $4$, respectively. The interface is shown in orange. The active region appears in light gray and can also be identified by the micro-vortices. 
A Movie is available in the Supplementary Information for $\activity = 3.5$, corresponding to (A.2). 
\textbf{B:} Starting from initial random noise, after spinodal decomposition the interface length decreases over time,
indicating coarsening of the structure. Depending on activity \activity, this coarsening is arrested,
with the gray boxes marking the corresponding dynamic equilibrium states. The averaged mean interface length in the arrested state increases approximately linearly with \activity (inlet).
\textbf{C:} Structure factor of the phase field $c$, normalized by the activity scale $K$
and its maximum value at that scale. At small $k$, the distribution flattens with increasing
\activity.
\textbf{D:} Velocity increment distributions in the active region show only a weak
dependence on \activity. The deviation from Gaussian statistics (gray dashed lines) is quantified by the kurtosis ${\cal K}$ (inlet).
\textbf{E:} Mean kinetic energy density in the active phase during coarsening. After the initial
build-up of the flow ($t<10$), the kinetic energy density increases until the system
reaches a dynamic equilibrium state, indicated by gray boxes. The averaged mean energy density in the dynamic equilibrium state
increases approximately linearly with \activity (inlet).
\textbf{F:} The energy spectrum peaks at values of $k$ smaller than the forcing scale $K$, indicating the presence of flow structures at larger scales. At small $k$, the
energy spectrum decreases linearly.
\textbf{G:} Probability distribution (pdf) for number of active (solid lines) and passive (dashed lines) domains in the dynamic equilibrium state. At high activity \activity,
the bicontinuous emulsion breaks up, resulting in the formation of active and passive
droplets, thus increasing the number of domains. The number of passive domains is significantly higher than that of active
domains (inlet).
}
\end{minipage}
\end{center}
\end{figure}

We first describe the Cahn-Hilliard-Generalized-Navier-Stokes (CHGNS)
equations, before computational results are tested against
various properties of active turbulence and two-phase systems. We thereby
mainly focus on parameter settings which lead to arrested coarsening and
explore the emerging dynamic equilibrium state, which is characterized by
internal boundary fluctuations leading to the formation of droplets and their
merging, reminiscent to active emulsions, see Figure \ref{fig:dynEq} and the
corresponding Movies in the Supplementary Information. Closer inspection of the geometry and topology of the two
regions within these states reveal asymmetries between the active and the
passive region. There are more passive than active droplets and the morphology of the active regions seems less fragmented but forms more elongated structures than the passive region. We quantify these asymmetries and explore their origin. Finally conclusions are drawn.

\section{Model}

We consider the GNS equations \cite{Slomkaetal_EPJSP_2015}, which relate to
the phenomenologically-proposed Toner-Tu-Swift-Hohenberg (TTSH) equations
\cite{Wensinketal_PNAS_2012,Dunkeletal_PRL_2013} for bacterial
suspensions. However, the TTSH equations, also considered in
\cite{Bratanovetal_PNAS_2015,Mukherjeeetal_PRL_2021,Mukherjeeetal_NP_2023}
neglect hydrodynamic interactions mediated by the solvent, which makes it a
dry system \cite{Alertetal_ARCMP_2022}. Combining the GNS equations with the
CH equations reads

\begin{align} 
\partial_t \uu + \uu \cdot \nabla \uu + \nabla p = \nabla \cdot \boldsymbol{\sigma} - c \nabla \mu, \quad \nabla \cdot \uu = 0 \label{eq:GNS}
\end{align}
with velocity $\uu(\mathbf{x},t)$, pressure $p(\mathbf{x},t)$, phase-field $c(\mathbf{x},t)$, chemical potential $\mu(\mathbf{x},t)$ and effective stress tensor $\boldsymbol{\sigma}(\mathbf{x},t)$ which comprises passive contributions from the intrinsic fluid viscosity and active contributions representing the forces exerted by the bacteria on the fluid. It reads $\sigma_{ij} = \nueff(\Lap,c) \left( \partial_i u_j + \partial_j u_i \right)$ with
\begin{align}
\nueff(\Lap,c) = \frac{c +1}{2} \nuact(\Lap) - \frac{c -1}{2} \nupas = \frac{c +1}{2} (\GO - \GT \Lap + \GF \Lap^2) - \frac{c -1}{2} \GO 
\end{align}
and $\Lap = \nabla \cdot \nabla$ the Laplacian, $\GO >0$, $\GT < 0$ and $\GF > 0$. These parameters are the simplest choice which lead to an isotropic active stress tensor $\boldsymbol{\sigma}$ \cite{Slomkaetal_PRF_2017}, which produces a band of unstable modes injecting energy to drive flows and produce patterns of vortices.  Making the definition of $\nueff$ to depend on $c$ allows to distinguish between the active regions ($c \approx 1$), the passive regions ($c \approx -1$) and the interface between them ($c \approx 0$). The equations are coupled to the CH equations to determine $c$ and $\mu$
\begin{align}
\partial_t c + \uu \cdot \nabla c = \nabla \cdot(M \nabla \mu), \quad \mu = \tilde{\sigma} \left(- \epsilon \Lap c + \frac{1}{\epsilon} W^\prime(c) \right)
\label{eq:CH}    
\end{align}
with rescaled interfacial tension $\tilde{\sigma}$, double-well potential $W(c) = \frac{1}{4}(1 - c^2)^2$, mobility $M$ and interface width $\epsilon$. In order to concentrate on the effect of the active region on the interface we consider the simplest possible model with equal density of the active and passive regions, which is here rescaled to $1$. For $\GT = \GF = 0$ eqs. \eqref{eq:GNS} - \eqref{eq:CH} reduce to the classical Cahn-Hilliard-Navier-Stokes (CHNS) equations, we therefore call the full model Cahn-Hilliard-Generalized-Navier-Stokes (CHGNS) equations.  We explore this model numerically using an approach similar to \cite{Linkmannetal_PRL_2019,Linkmannetal_PRE_2020,Linkmannetal_JFM_2020} and approximate the effective viscosity in the active region $\nuact$ by a piecewise constant function in reciprocal space. 
We define
$\nuact(k < k_0) = \tilde{\Gamma}_0$, $\nuact(k_0 \le k \le k_1) = \tilde{\Gamma}_2$ and $\nuact(k_1 < k) = \tilde{\Gamma}_4$,
with $\tilde{\Gamma}_0 > 0$, $\tilde{\Gamma}_2 < 0$ and $\tilde{\Gamma}_4 > 0$, with $k$ being the wavenumber. $\tGO$ describes the damping of long-wavelength perturbations on scales much larger than the typical correlation length of the coherent flow structures, whereas $\tGT$ and $\tGF$ account for the growth and damping of modes at intermediate and small scales. The considered approximation covers the essential features of the full model \cite{Linkmannetal_PRL_2019,Backofenetal_PNAS_2024}. The intermediate scales at which energy is injected are within the activity band $k_0 \le k \le k_1$. We consider $k_0 = 33$ and $k_1 = 40$, chosen to be consistent with \cite{Backofenetal_PNAS_2024}. This determines the activity scale $K = (33 + 40)/2 = 36.5$. We keep $\tilde{\Gamma}_0=1.1\cdot10^{-3}$ and $\tilde{\Gamma}_4 = 1.1\cdot10^{-2}$ and only vary $\tilde{\Gamma}_2$. In the following we use $\activity := -\tilde{\Gamma}_2 / \tilde{\Gamma}_0$ as the activity parameter of the model. We consider $\tilde{\sigma} = 10^2$. Another modeling assumption considers the mobility, which is chosen to be constant $M = 0.01$, following \cite{MPCMC_JFM_2013}. For alternative approaches see \cite{A_ARMA_2009}. Based on the analysis in these works for the CHNS equations the limit $\epsilon \to 0$ of the CHGNS equations can be expected to lead to a two-phase system with one bulk phase ($c = -1$) governed by the NS equations, one bulk phase ($c = 1$) governed by the GNS equations and the interface conditions $[\mathbf{u}] = 0$ and $[-p \mathbf{I} + \boldsymbol{\sigma}] \cdot \mathbf{n} = \sigma \kappa \mathbf{n}$, with $[\cdot]$ denoting the jump across the interface, $\mathbf{I}$ the identity matrix, $\mathbf{n}$ the normal to the interface, $\sigma$ the interfacial tension and $\kappa$ its mean curvature. We do not explore this limit and only consider a constant interface width with $\epsilon = 0.1$. Furthermore we consider only a symmetric two-phase system with $\int_\Omega c \; d\mathbf{x} = 0$, with $\Omega$ the computational domain. This is chosen as a rectangle in 2D with periodic boundary conditions, which allows for efficient numerical solution using a vorticity-stream-function formulation of the equations and a spectral method, adapting the approach for the GNS equations in \cite{Backofenetal_PNAS_2024} and 
solution methods for the NSCH equations \cite{padhan2025cahn}, see Supplementary Informations for details.

\section{Results}
\subsection{Activity arrested coarsening and flow structures}

Within the considered parameter range the evolution of the CHGNS equations, if started from initial random noise around $c = 0$, is governed by spinodal decomposition and subsequent coarsening, similar to the classical CHNS equations. However, due to the active character of one region energy is constantly injected into the system. This has an effect on the coarsening process, which, depending on the strength of activity, gets arrested and leads to dynamic equilibrium states. We are mainly interested in these dynamic equilibrium states, which can be considered as a dynamic emulsion. Line-integral-convolution (LIC) visualizations illustrate snapshots of these states at different activity levels, see Figure~\ref{fig:dynEq}~A.0–A.3 with increasing activity from left to right. At small activity ($\activity = 2.5$), macroscopic phase separation occurs on the scale of the simulation domain, with interface fluctuations. With increasing activity ($\activity = 3.0, 3.5, 4.0$), the 
fluctuation intensity grows and the system develops continuously fluctuating 
active and passive domains with a characteristic length scale smaller than 
the system size. The characteristic length scale decreases with increasing $\activity$. The 
resulting states form active, dominantly bicontinuous emulsions, with 
occasional droplets that emerge from pinch-off events of protrusions of the interface, see the corresponding Movies in the Supplementary Information.
The droplets are transported and gradually vanish due to coalescence or Ostwald ripening. Within the active region the active stress continuously drives the flow, forming micro-vortices at the length scale set by the activity band. With sufficiently large active stress, the flow self-organizes into larger coherent structures, including jets and mesoscopic vortices, in which the mean flow rotates as a whole, see the corresponding Movies in the Supplementary Information. Interface fluctuations occur on scales larger than those of the micro-vortices in the dynamic equilibrium state. The flow in the passive region primarily follows the interface, and the velocity field smooths the fluctuations induced by the interface. 

To understand how the system approaches its dynamic equilibrium states, we analyze
both the evolution of the interface length and the development of the flow
field. In Figure~\ref{fig:dynEq} B the evolution of the 
total interface length is shown for different activity levels.
After an initial phase and before coarsening arrest the interface length decreases faster than the $t^{-1/3}$ scaling of the CH equations \cite{voorhees1992ostwald}. It approaches the $t^{-1/2}$ 
scaling of the CHNS model \cite{padhan2025cahn} in the late stages for low activity ($\activity = 2.5$). 
For higher activity, coarsening arrests at a characteristic 
length scale, and the interface length fluctuates around a constant mean, i.e., a dynamic equilibrium, which is indicated by the gray boxes. The mean interface length in the dynamic equilibrium states increases approximately linearly with activity (Figure~\ref{fig:dynEq} B inlet). The evolution of the kinetic energy with time is shown in Figure~\ref{fig:dynEq} E. 
Starting from small initial velocity fluctuations, the flow field builds up and exhibits a characteristic overshoot at $t \approx 2$. For $t \gtrsim  10$ the kinetic energy enters the same long-time regime as the interface length, both quantities evolve in parallel, with decreases in interface length accompanied by increases in kinetic energy. Within the dynamic equilibrium states the kinetic energy fluctuates around a well-defined mean, again indicated by the gray boxes. This mean value increases approximately linearly with activity (Figure~\ref{fig:dynEq} E inlet). Thus, larger \activity lead to higher kinetic energies and larger interface length and therefore smaller morphological structures in the dynamic equilibrium state. 

We next analyze the morphological structure of the dynamic equilibrium states. Therefore all quantities are averaged in time, over the intervals $T$ indicated by the gray boxes in Figure~\ref{fig:dynEq} B and E. The mean radial distribution of the phase field structure factor density, $\avT{S(k)}$, is shown in Figure~\ref{fig:dynEq} C. At low activity ($\activity = 2.5$), the system separates on the scale of the domain, forming a single compact active region (Figure~\ref{fig:dynEq} A.0). Accordingly, $\avT{S(k)}$ increases monotonically towards smaller $k$, corresponding to larger length scales. It is maximal at the smallest wavenumber. The structure of the phase field can be quantified as anti-hyperuniform. For arrested coarsening, corresponding to larger activities, this behavior changes. $\avT{S(k)}$ gradually flattens at small $k$, and for sufficiently large activity ($\activity = 3.5, 4.0$) it becomes essentially horizontal in this region, i.e., $\avT{S(k)} \propto k^{\beta}$ with $\beta \approx 0$, which is termed non-hyperuniform. The extent of this flat region grows with increasing activity, reflecting a breakdown of large-scale order. Thus, in the dynamic equilibrium states, the phase field morphology is not hyperuniform, in contrast to the self-similar structures observed in classical CH or CHNS coarsening, for which $\avT{S(k)} \to 0$ as $k \to 0$ \cite{Maetal_JAP_2017,padhan2025suppression}. The structure is therefore also different to other active extensions of the CH equations, such as active Model B \cite{wittkowski2014scalar} or active Model B+ \cite{Nardini2017activeB+}, which show characteristics of hyperuniformity during coarsening \cite{zheng2024universal,de2024hyperuniformity}, but similar to active Model H \cite{PhysRevLett.115.188302,Tjhung2018active}, which leads to arrested coarsening for contractile forces with non-hyperuniform morphologies \cite{padhan2025suppression}.

The radial distribution of the kinetic energy, also averaged over time within the dynamic equilibrium state, $\avT{E(k)}$, is shown in Figure~\ref{fig:dynEq} F. 
Around the activity band, $k \approx K$ indicated by the gray shaded region, $\avT{E(k)}$ exhibits a local maximum, corresponding to energy injection by the active stress. 
For larger $k$, corresponding to smaller length scales, $\avT{E(k)}$ rapidly decreases due to viscous dissipation and small scale damping. For smaller $k$, corresponding to larger length scales, $\avT{E(k)}$ initially drops sharply, then increases approximately linearly, reaching a global maximum, and for even smaller $k$ decreases with $\avT{E(k)} \propto k^\beta$ with $\beta \approx 1$. As this relates to a scaling with $\beta \approx 2$ for the angular-averaged spectral density of the vorticity autocovariance, see \cite{Backofenetal_PNAS_2024}, this indicates characteristics of hyperuniformity of the vorticity field, as for the (one-phase) GNS equations before condensation and various experimental systems, including bacterial suspension, sperm
suspension, self-propelled Janus particles and tissue cell monolayers \cite{Backofenetal_PNAS_2024,Alertetal_ARCMP_2022}. This behavior is largely independent of activity. In contrast, the slope and amplitude of the increase from the activity band to the global maximum 
are activity-dependent, becoming steeper and larger for higher
$\activity$, respectively. Although the scaling does
not reach the theoretical $-5/3$ slope expected for a full inverse cascade, this trend indicates partial inverse energy transfer from the activity band toward larger scales. Consequently, flow structures larger than the activity scale emerge; in biological terms, the flow self-organizes into larger structures. As activity increases ($\activity > 2.5$), the wavenumber corresponding to the maximum of $\avT{E(k)}$ shifts to larger $k$, i.e., smaller structures. We will demonstrate that these structures are connected to the morphology of the phase field. At lower activity, see Supplementary Information, the inverse cascade eventually vanishes, and no self-organization of flow structures is observed.  

In order to further support the observed characteristics of turbulence, we show in Figure \ref{fig:dynEq}~D the probability distribution function (pdf) of the velocity increments in the active region, 
averaged in the dynamic equilibrium state. The pdf deviates from a perfect Gaussian at larger increments, developing heavier tails with increasing activity. This deviation is quantified by the kurtosis ${\cal K}$ (Figure \ref{fig:dynEq}~D inlet), 
which rises from approximately 3.25 to 3.4 as $\activity$ increases. Such elevated kurtosis values indicate intermittency 
and weakly turbulent flow. The flow field in the active region thus shows similar properties as the flow field in the (one-phase) GNS equations \cite{Backofenetal_PNAS_2024}. The results for the kinetic energy and the velocity increments also compare well with the two-phase turbulent flow of two turbulent phases studied in~\cite{perlekar2017two}. 
There, turbulent stirring arrests coarsening in a manner that depends 
on the stirring strength, and the inverse cascade is similarly bounded 
by a characteristic length scale. However, \cite{perlekar2017two} does not investigate any direct connection between 
the characteristic length scale of the fluid flow  and the structural length scale of the phase field.

We now address this connection. With increasing activity ($\activity > 2.5$), the characteristic length scale of the dynamic emulsion decreases, and progressively more small active and passive droplets appear. We define such droplets as compact regions of passive material surrounded by active material and vice versa. While already observable in Figure~\ref{fig:dynEq}~A.1--A.3, this property is quantified in
Figure \ref{fig:dynEq}~G, by showing the probability distribution of active (solid lines) and passive (dashed lines) droplets in the dynamic equilibrium states. As $\activity$ increases, 
the number of droplets grows, and there are consistently more passive droplets 
than active ones (Figure \ref{fig:dynEq}~G inlet). In the following, we connect this observed asymmetry to an asymmetry between the flow fields in the active and passive regions, in particular focusing on how the flow structure within the active region depends on the morphology of the phase field.

\subsection{Asymmetry between active and passive regions}


We propose the following mechanism for the observed asymmetry between active 
and passive droplets: Droplet formation occurs through pinch-off events triggered by interface fluctuations, whose strength is controlled by the local active flow structure. Larger available space in the active region allows the formation of larger 
vortices with higher energies, which in turn generate stronger interface fluctuations. 
For active protrusions, locally available space is small, suppressing 
vortical activity and reducing fluctuation amplitudes, which makes pinch-off 
less likely. Passive protrusions, however, are surrounded by active region with potentially larger available space, leading to stronger fluctuations that promote pinch-off. This asymmetry in fluctuation strength naturally results in more passive than 
active droplets.

\begin{figure}[htb]
  \noindent
  \begin{tabular}{ccc}
     A \includegraphics*[height=3.7 cm]{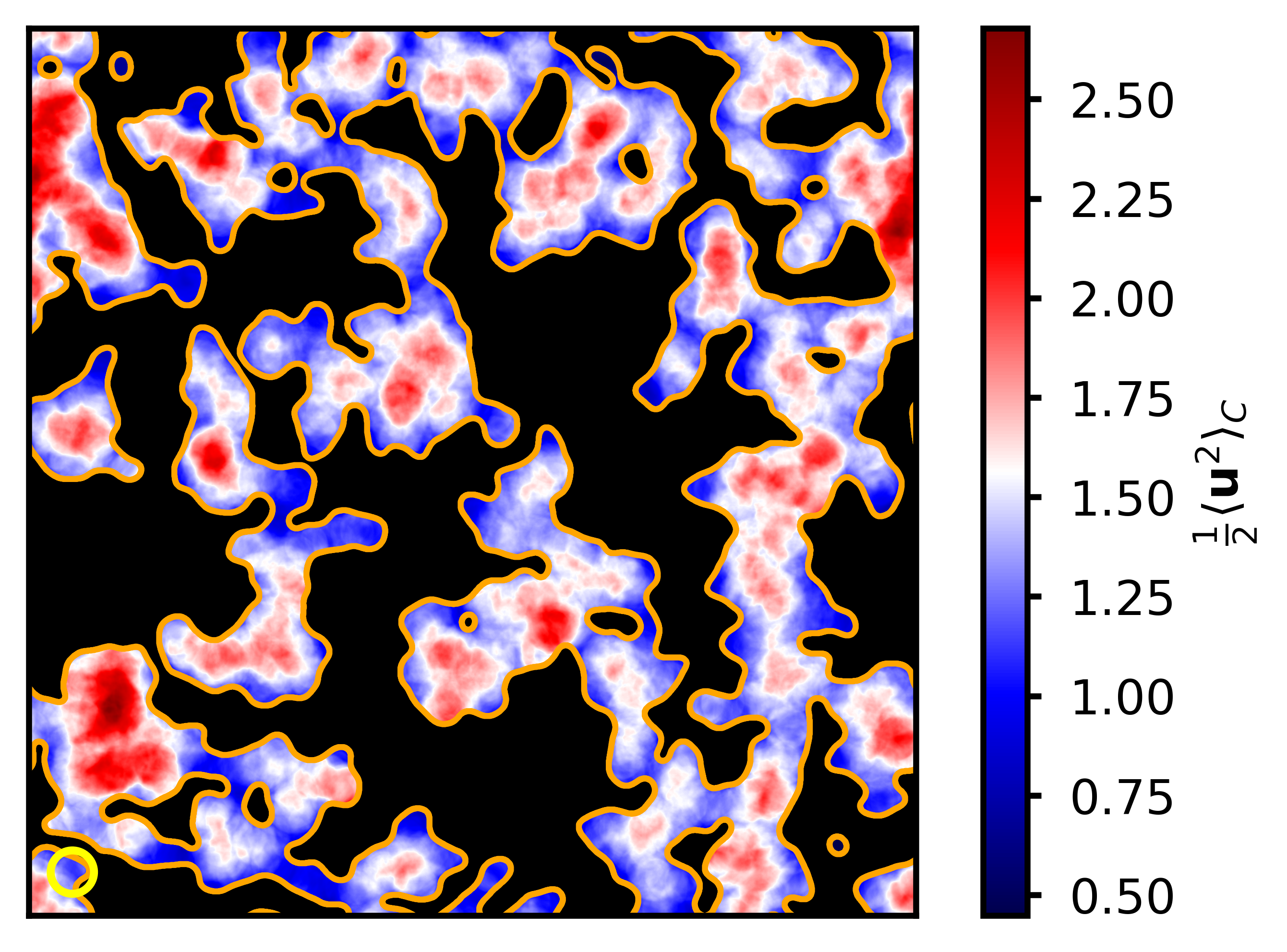}
    &B\includegraphics*[height= 3.7 cm]{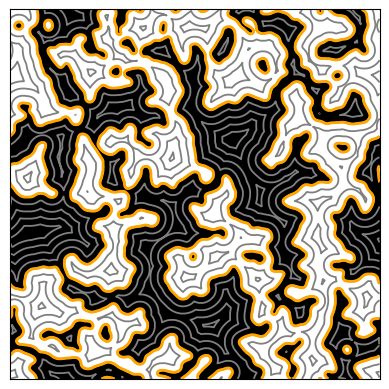} 
    &C \includegraphics*[width=0.3\textwidth]{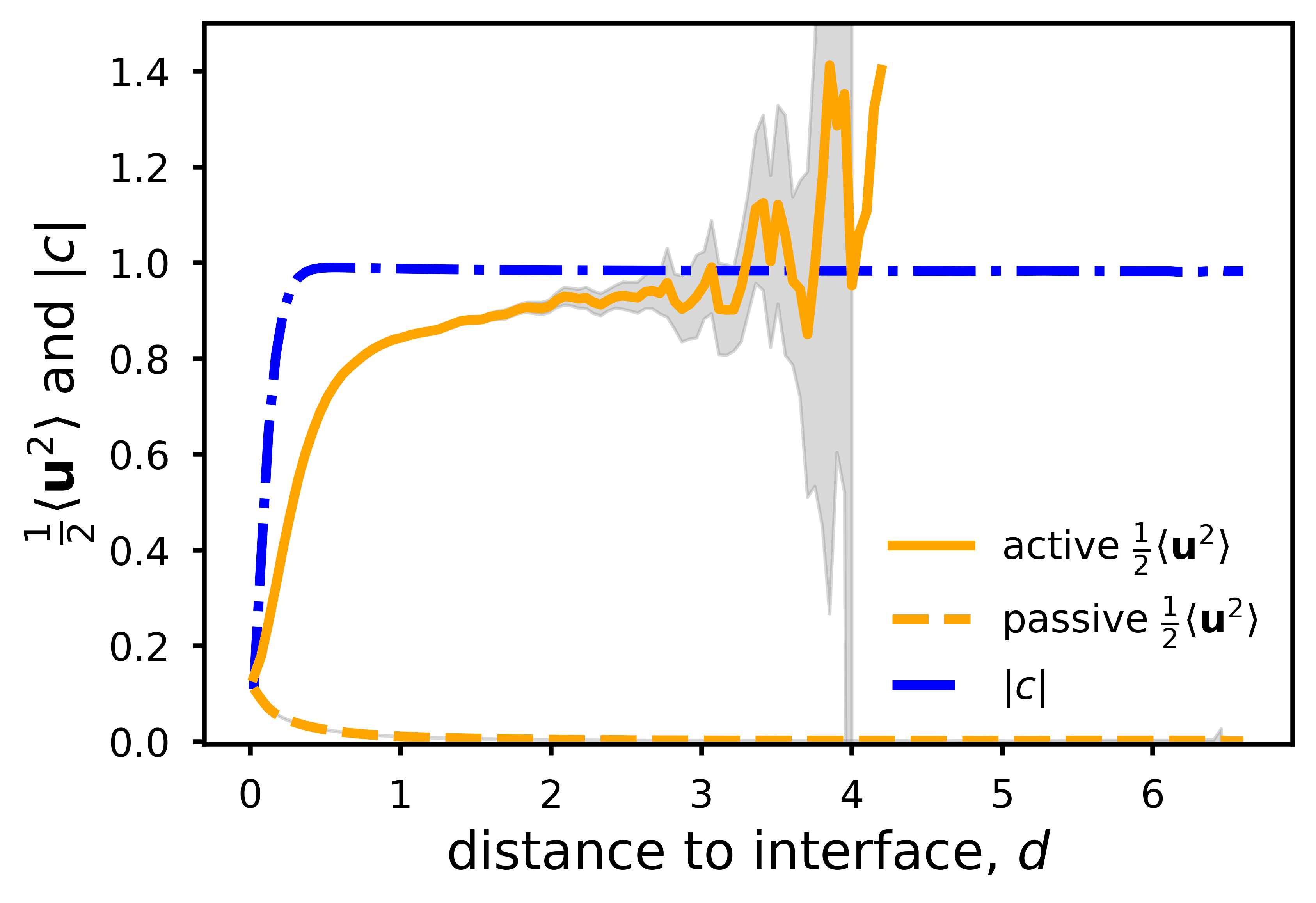} 
       \end{tabular}
  \begin{center}
    \begin{minipage}{0.95\textwidth}
      \caption[short figure description]{
        \textbf{A:} Local averaged mean kinetic energy $\frac{1}{2}\avC{\mathbf{u}^2}$ in the active region. The
$\mathbf{u}^2$ field is averaged inside the active region within a circle of radius
$r = 1.23$ (yellow circle). Highest energies are mostly found deep inside the active region.
Narrow protrusions and bridges tend to exhibit lower kinetic energy. In
general, $\frac{1}{2}\avC{\mathbf{u}^2}$ decreases toward the interface. The passive region is
shown in black and the interface in orange.
\textbf{B:} Isolines of the distance to the interface $d$. The passive region is shown in black and the interface in orange. 
\textbf{C:} Mean kinetic energy density $\frac{1}{2}\av{\mathbf{u}^2}$ (orange) and phase field $c$ (blue) as functions of the distance to the interface $d$. In the active region, $\frac{1}{2}\av{\mathbf{u}^2}$ increases sharply for $d < 1$, forming a boundary layer. In the bulk region ($d > 1$), $\frac{1}{2}\av{\mathbf{u}^2}$ further increases with increasing $d$, but much more moderately. In the passive region, $\frac{1}{2}\av{\mathbf{u}^2}$ is significantly smaller than in the active region and decreases with increasing $d$. The interface width of the phase field is much smaller than the thickness of the fluid boundary layer. The fluctuations and increasing error bars for large $d$ for the active region result from limited available data.
        \label{fig:asym} \label{fig:vvvsd}
        }
\end{minipage}
\end{center}
\end{figure}

In order to support this proposed mechanism we analyze the differences in the flow structure of the active and passive regions in the dynamic equilibrium state, focusing on the case $\activity = 3.5$. We show that kinetic energy is primarily concentrated in the active region, that larger available space in the active region allows the formation of more energetic flow structures, and that these stronger flows in turn generate enhanced interface fluctuations. 

We begin by analyzing the mean kinetic energy locally within the active region. This quantity is shown in Figure~\ref{fig:vvvsd}~A for a single time instance. Within the active region, the kinetic energy is averaged over a circle of
radius $r = 1.23$. Regions with the highest kinetic energy are predominantly
found in the largest connected domains of the active region. In contrast,
protrusions or thin bridges of active material connecting larger regions tend to
exhibit weaker, less energetic flow. To quantify this observation, we introduce the distance to the
interface $d$ in Figure~\ref{fig:vvvsd} B. The isolines of constant distance to the interface highlight a pronounced asymmetry in the morphological structure. Although the passive region contains a larger number of small droplets, it also exhibits
the regions with the largest distances to the interface. While this is observable in the considered snapshot, it also results from the statistical analysis, with the largest passive regions showing on average distances to the interface $d$ which are roughly 1.5 times the largest distances within the active region.
In Figure~\ref{fig:vvvsd}~C the mean kinetic-energy density is shown as a function of the distance to the interface $d$, separately for the active (solid line) and passive 
(dashed line) regions, together with the phase field $c$. At a given distance $d$, the kinetic energy at this distance
is averaged in the whole simulation domain and over the set of time frames in the time interval $T$ in
the dynamic equilibrium state. In the passive region, the kinetic-energy density decreases rapidly with $d$. In contrast, in the active region the kinetic-energy density 
increases with $d$, first ($d < 1$) rapidly, forming a fluid boundary layer, and later ($d > 1$), more moderately. For large distances 
($d > 3$), however, statistical uncertainties dominate, leading to large 
fluctuations and wide confidence intervals. This is due to the limited number 
of large active domains. Only very few exhibit distances 
greater than $3$, which results in small sample sizes. 
\begin{figure}[htb]
  \noindent
  \begin{tabular}{ccc}
     A \raisebox{-0.03cm}{\includegraphics*[height=3.73 cm]{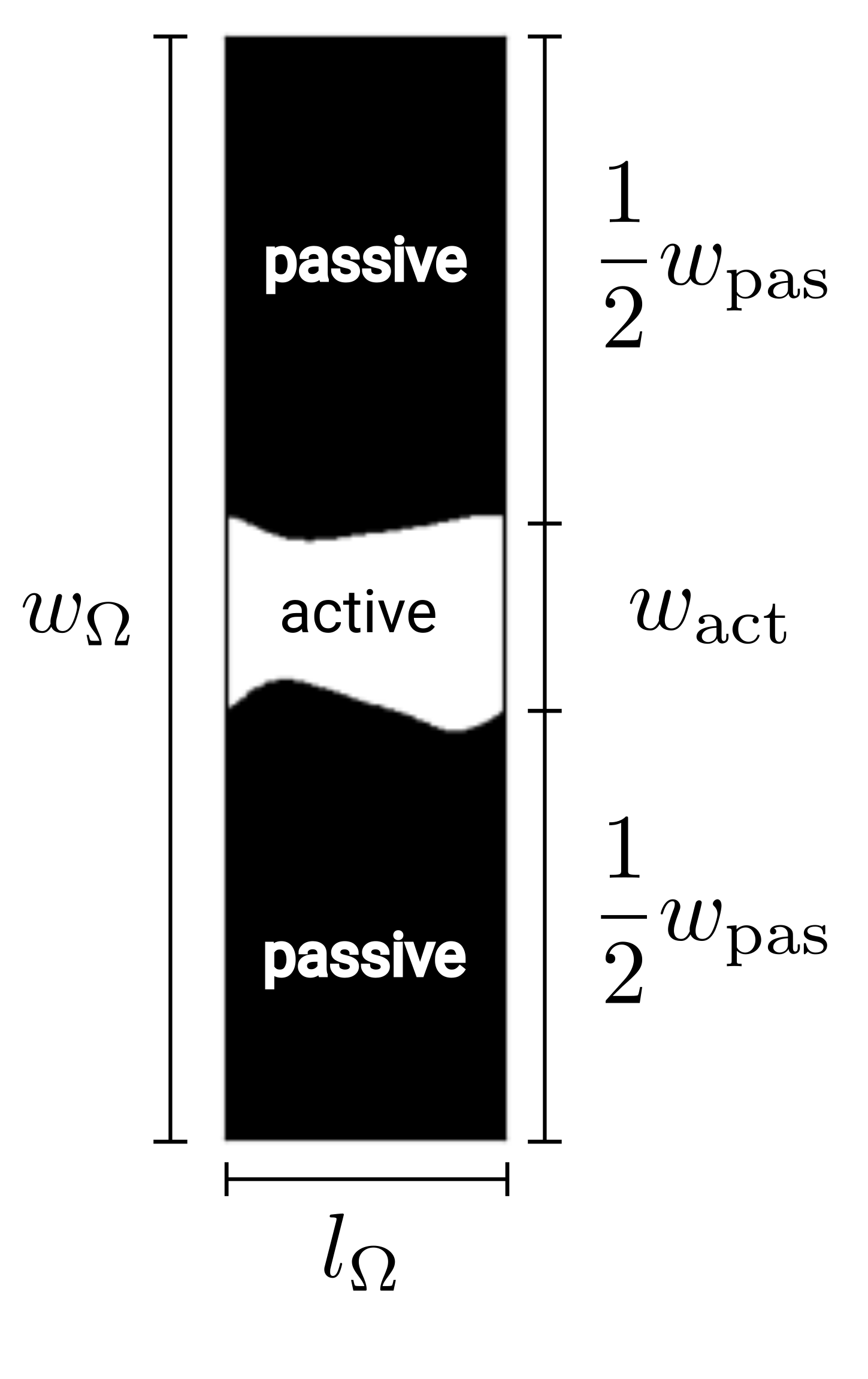}}
    &B \includegraphics*[width=0.3\textwidth]{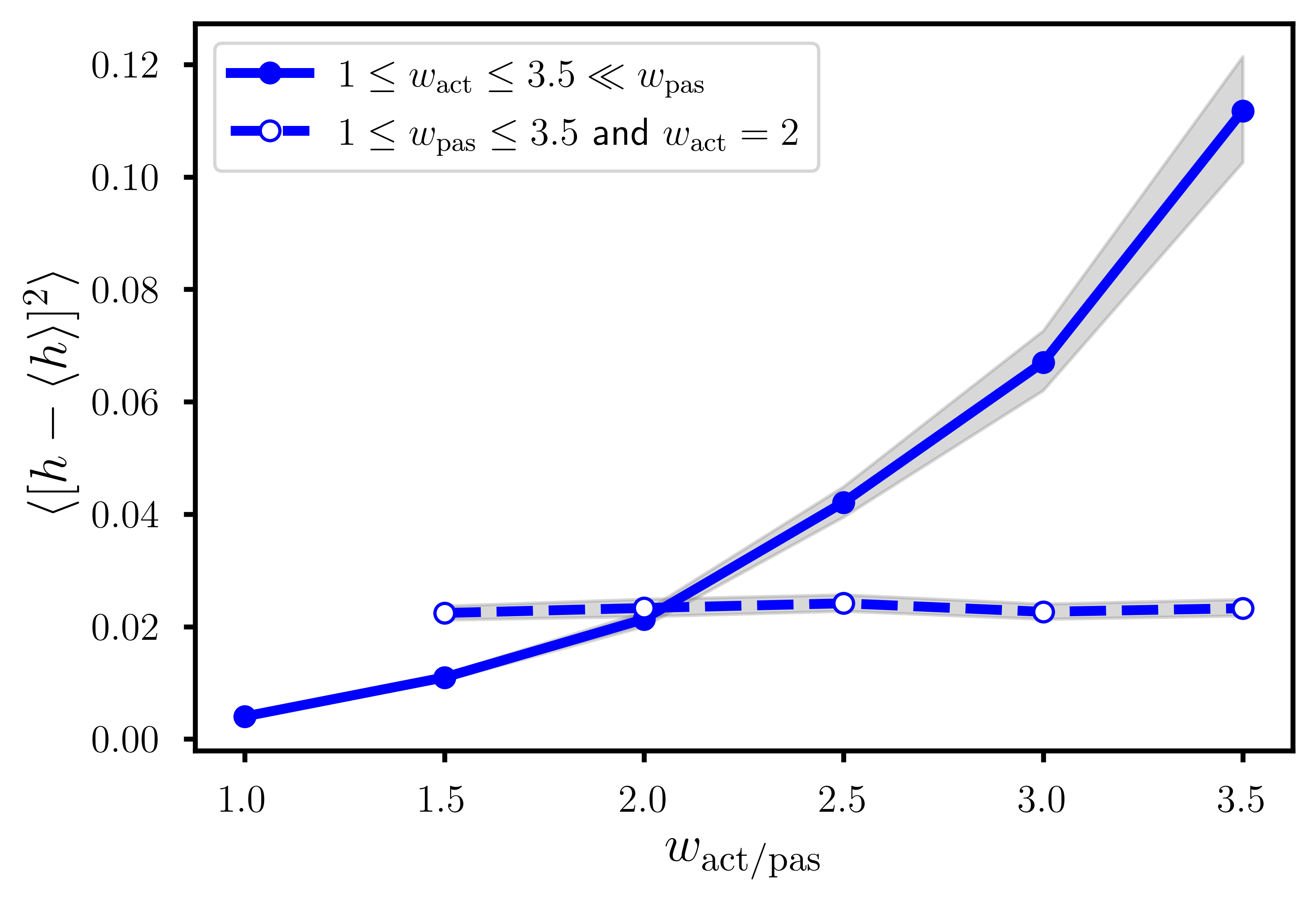}
    &C \raisebox{0.02cm}{\includegraphics*[width=0.3\textwidth]{slabs2}}
  \end{tabular}
  \begin{center}
    \begin{minipage}{0.95\textwidth}
    \caption[short figure description]{ Interface height fluctuation of a phase separated situation in a thin rectangular domain within dynamic equilibrium states. \textbf{A:} Geometric setting. \textbf{B:} Interface height fluctuation $\langle
      [ h - \langle h \rangle ]^2 \rangle$ for two different settings with varying height $w_{\rm act}$ and $w_{\rm pas}$, respectively. Closed symbols: $1 \leq
        w_{\rm act} \leq 3.5$ and $w_{\rm pas} \gg 3.5$. Open symbols: $w_{\rm act} = 2$ and $1.5 < w_{\rm pas}< 3.5$). No fluctuation is measured for $w_{\rm pas} = 1$, as in this case the interface becomes unstable and a passive droplet forms. For both cases $l_\Omega=\pi$. \textbf{C:} LIC visualization of the flow field and the interfaces (orange lines) at a time instance within the dynamic equilibrium states for various $w_{\rm act}$ (corresponding to closed symbols in \textbf{B}, demonstrating the fluctuation increase with increasing $w_{\rm act}$. The corresponding flow fields to the open symbols in \textbf{B} are not shown as there is no affect on the interface fluctuation for increasing $w_{\rm pas}$. 
        \label{fig:slabfluc}
        }
\end{minipage}
\end{center}
\end{figure}

The increase of the mean kinetic-energy density at larger distances in the 
active region suggests that the flow intensity depends on the local size of the active region. To examine this relation more directly and to isolate its influence on interface fluctuations, we study a simplified situation consisting of one active and one passive region in a thin rectangular domain, as illustrated in Figure~\ref{fig:slabfluc}~A.
First, we initialize an active region of height $w_{\rm act}$ within a large passive region of height $w_{\rm pas}$. We vary $1 \leq w_{\rm act} \leq 3.5$ and adjust $w_{pas}$ such that $w_{\rm pas} \gg w_{\rm act}$, e.g., exploring the impact of an increase in active region in a passive bath. The result is shown in Figure~\ref{fig:slabfluc}~C, which uses LIC visualization of the flow field and highlights the fluctuations of the interfaces at a time instance within the dynamic equilibrium states. The corresponding interface height fluctuations $\langle[ h - \langle h \rangle ]^2 \rangle$ are shown by the closed symbols and the solid line in Figure~\ref{fig:slabfluc}~B, which indicate a polynomial increase with increase of $w_{\rm act}$. Second, we consider an active region of fixed height $w_{\rm act} = 2$ and vary $1 \leq w_{\rm pas} \leq 3.5$, e.g. exploring the impact of an increase in passive region. The corresponding flow fields are not shown, only the interface height fluctuations are plotted in Figure~\ref{fig:slabfluc}~B using open symbols and the dashed line, which is independent of $w_{\rm pas}$. Overall this behavior indicates that larger active regions generate stronger flow fluctuations and therefore inject more energy into the interface dynamics.
At smaller activities, see Supplementary Information, an increase in fluctuations which saturates at a characteristic distance can be observed, or, for even smaller activities with no inverse cascade, the interface fluctuations become independent of the size of the active region. 

\begin{figure}[htb]
  \noindent
  \begin{tabular}{cc}
    A \includegraphics*[height = 4 cm]{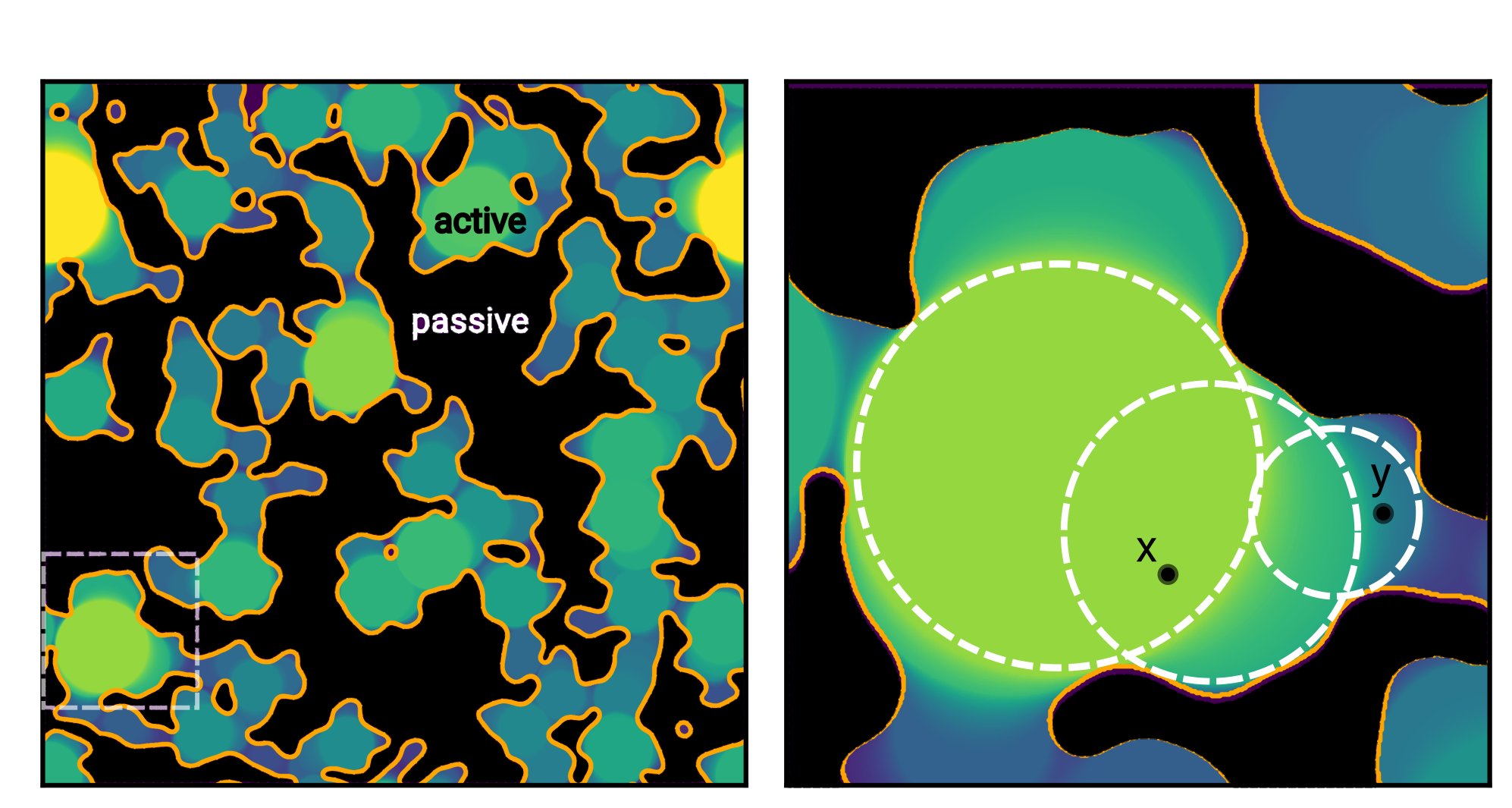}
    &B \includegraphics*[width=0.3\textwidth]{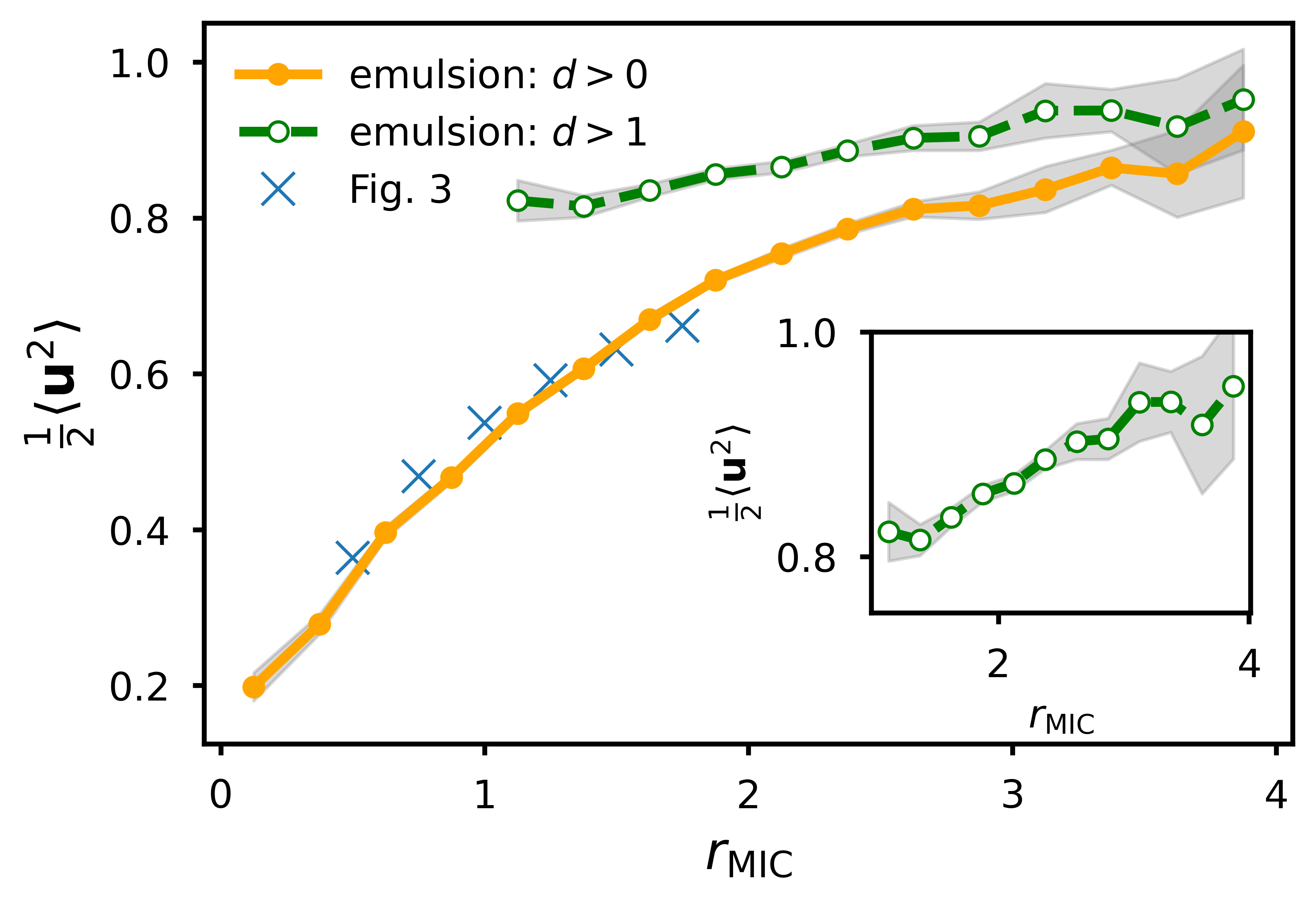}
    \end{tabular}
  \begin{center}
    \begin{minipage}{0.95\textwidth}
      \caption[short figure description]{
\textbf{A:} Definition of the available space for fluid organization using
maximum inscribed circles (MIC). Three inscribed circles are shown in the closeup. The
available space at a point is measured by the radius of the MIC, \rmic, which is color coded.
At $\mathbf{x}$, the available space is defined by the largest circle, while at
$\mathbf{y}$, it is defined by the smallest circle shown. \textbf{B:} Mean kinetic energy as a function of \rmic. The kinetic energy is averaged over the entire
region ($d > 0$, orange), as well as  excluding the boundary layer ($d > 1$, green).
Crosses show mean kinetic energies for the simplified setting from Figure~\ref{fig:slabfluc} (closed symbols), with $\rmic \approx 0.5 w_{\rm act}$. The inlet considers a different scaling to highlight the increase of the mean kinetic energy for larger \rmic, independent of the boundary layer.
        \label{fig:rmic}
        }
\end{minipage}
\end{center}
\end{figure}

While these results support our proposed mechanism within the simplified setting, we next need to quantify the found buildup of kinetic energy in larger active regions of the dynamic emulsion. Therefore, we introduce a geometric measure for the locally available space for active fluid motion. In the active region, we consider the set of all circles that lie entirely within the active region (the set of inscribed circles). The available space at a point~$\mathbf{x}$ is then defined by the area of the maximal inscribed circle (MIC) that contains $\mathbf{x}$, see Figure \ref{fig:rmic}~A. The area of the MIC provides an upper bound for the size of flow structures that can develop locally. 

In Figure~\ref{fig:rmic}~B, the mean kinetic energy is plotted against the radius of the MIC,
~\rmic. Again, as for the distance to the interface, for a given~\rmic, the kinetic energy within this~\rmic is averaged in the whole domain and over a set of time frames in the time interval $T$ in the dynamic equilibrium state. This illustrates how the local mean kinetic energy of the
fluid depends on the space available for organizing larger flow
structures. When evaluating the entire active phase in the dynamic emulsion
(orange), the mean kinetic energy increases with~\rmic, though the rate
of increase slows for larger values. The same analysis applied to the simplified setting in Figure~\ref{fig:rmic}~B with varying $w_{\rm act}$ (crosses) and $\rmic \approx 0.5 w_{\rm act}$,  essentially follows the curve, indicating that the flow properties in the dynamic emulsion and in the simplified setting are comparable. However, part of this increase in kinetic energy is attributable to boundary layer effects in the flow, as visible in Figure~\ref{fig:asym}~C. 
To separate the boundary-layer contribution from a potential bulk effect due to flow organization, we additionally show the mean kinetic energy as a function of~\rmic evaluated only in regions outside the boundary
layer ($d > 1$), Figure~\ref{fig:rmic}~B (green). In this case, the mean
kinetic energy is larger than in the full-field average, as the
low-energy boundary layer is excluded from the averaging process. Even under this restriction, the mean kinetic energy still increases
with~\rmic. Although the trend is less pronounced, it reflects a genuine
bulk effect. Notably, there remains an increase of about
20\% when \rmic\ grows from~2 to~4. This supports the conclusion that
the mean kinetic energy depends on the available space for organizing
larger flow structures.

\begin{figure}[htb]
  \noindent
  \begin{tabular}{cc}
     \includegraphics*[width=0.3\textwidth]{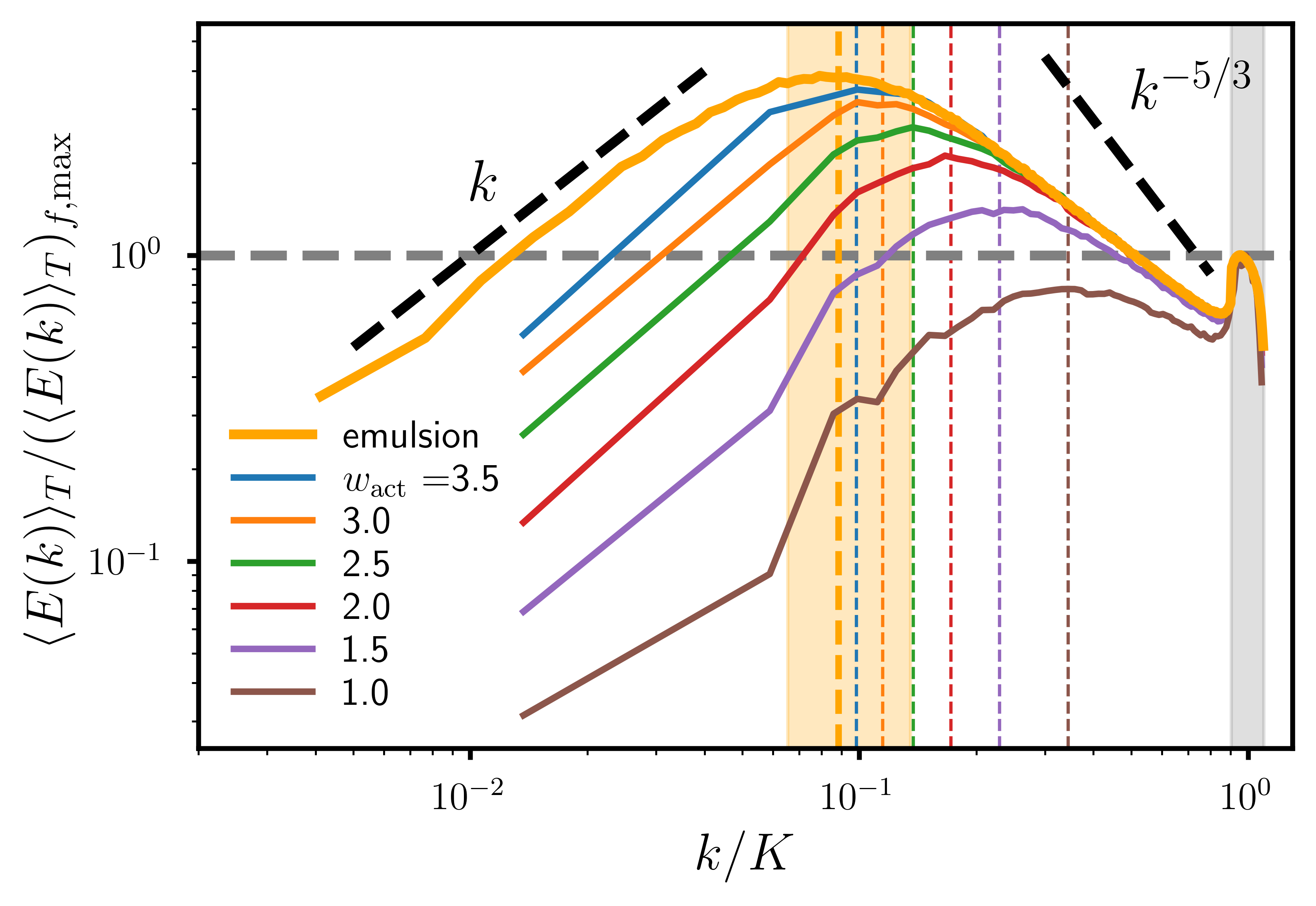} & \\  
    \end{tabular}
  \begin{center}
    \begin{minipage}{0.95\textwidth}
      \caption[short figure description]{
        Comparison of fluid structure, \avT{E(k)}$_{\rm max}$, with the available space $\av{\rmic}$. 
The mean energy spectrum is shown for the emulsion (thick line) and for single
slabs with width $w$
(thin lines, see Fig.~\ref{fig:slabfluc}). Horizontal lines indicate the length scale defined by 
the available space, $k = 2 \pi / \rmic$, for each setup. As slab width $\rmic$ increases, 
the maximum of $E(k)$ shifts to larger length scales and grows in magnitude, 
closely following the available space. For the emulsion, the mean available space 
$\av{\rmic}$ is shown as a thick line, with light orange shading representing the standard deviation.
\label{fig:struct}
        }
\end{minipage}
\end{center}
\end{figure}

Figure~\ref{fig:struct} links the length scale defined by the available space~\rmic to the flow structure expressed through the energy spectrum~$\avT{E(k)}$, which is shown for the dynamic emulsion (thick line) and for the simplified setting from Figure~\ref{fig:slabfluc} (closed symbols) (thin lines). The vertical lines mark the corresponding available space via $k = 2\pi / \rmic$. For the simplified setting, $\rmic \approx 0.5 w_{\rm act}$, whereas in the emulsion we use the mean available space \av{\rmic} measured in the dynamic equilibrium state. In all cases,
\av{\rmic} aligns well with the position of the maximum of $\avT{E(k)}$,
indicating that the flow structure adapts, on average, to the local
geometry. This shows that the flow in the active region self-organizes into the largest possible vortex permitted by the available space. As larger vortices correspond to larger flow intensity and thus stronger interface fluctuations and more frequent pinch-offs, this explains the asymmetry in the number of active and passive droplets, as similar effects are not present in the passive phase.

\begin{figure}[htb]
  \noindent
  \begin{tabular}{ccc}   
    A \includegraphics*[width=0.3\textwidth]   {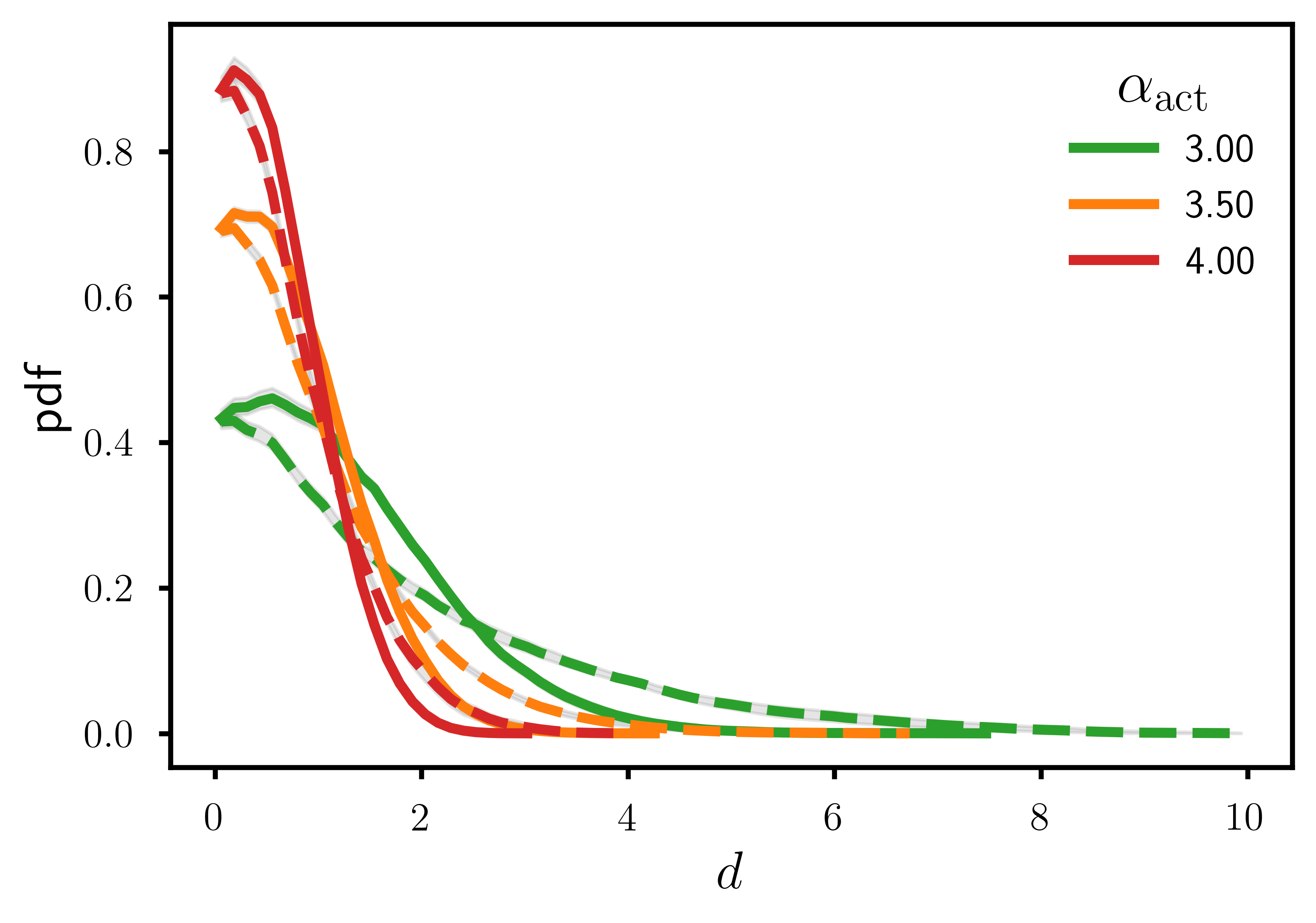}
    &B \includegraphics*[width=0.3\textwidth]{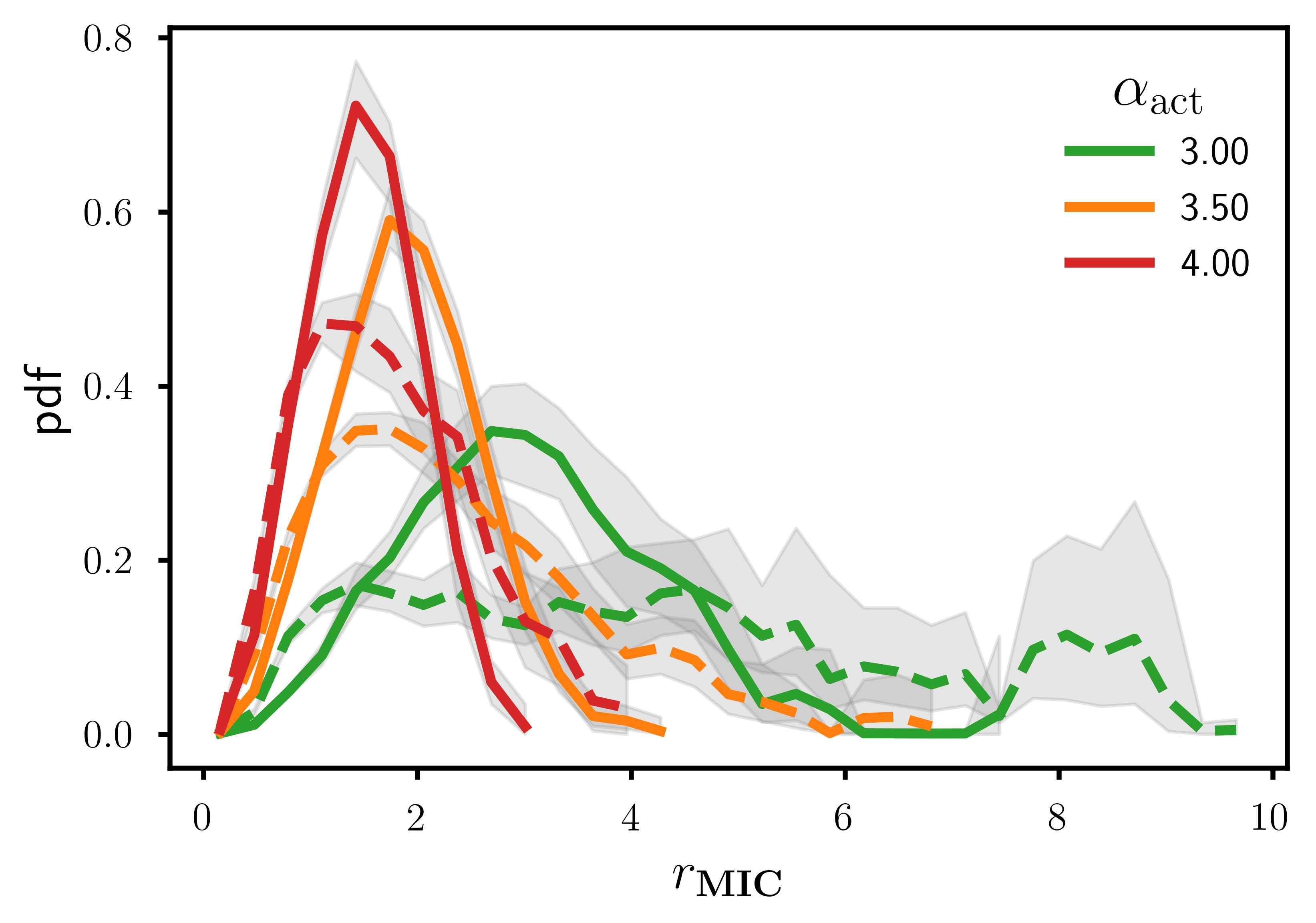}
    &C \includegraphics*[width=0.3\textwidth]{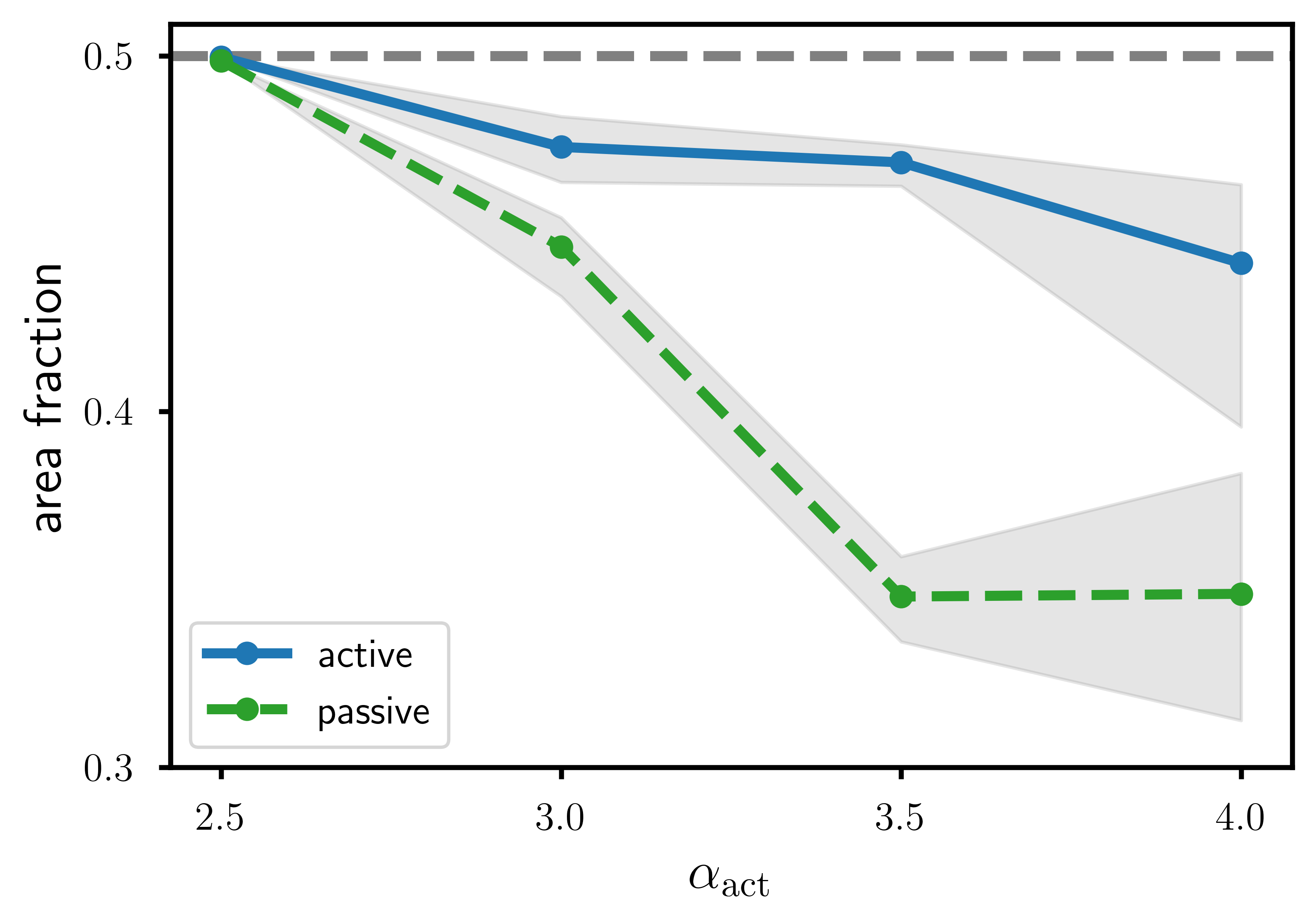}
      \end{tabular}
  \begin{center}
    \begin{minipage}{0.95\textwidth}
      \caption[short figure description]{ \textbf{A:} Probability distribution (pdf) for distance to interface $d$ in active (full line) and passive (dashed line) regions for $\activity = 3.0, 3.5$ and $4.0$. \textbf{B:} Probability distribution (pdf) for the radius of the MIC \rmic in active (full line) and passive (dashed line) regions for $\activity = 3.0, 3.5$ and $4.0$. 
      \textbf{C:} Area fraction of the largest connected domain for active (full line, blue) and passive (dashed line, green) within the total domain as a function of \activity.
\label{fig:struct2}
        }
\end{minipage}
\end{center}
\end{figure}

At last we analyze the morphologies of the active and passive regions. We consider the distance function to the interface $d$, the radius of the MIC $\rmic$, now evaluated in the active and the passive regions, and the area fractions of the largest connected domains in the active and passive regions for different activities \activity. They are shown in Figure \ref{fig:struct2} A - C, respectively. In Figure \ref{fig:struct2} A and B the probability distribution functions (pdf) for the dynamic equilibrium states are shown. They  indicate an asymmetry between the active and passive regions. With respect to $d$ this becomes apparent in the longer tails of the passive region, which is in agreement with the results discussed in Figure \ref{fig:asym} B and C. But also for small $d$ the curves differ. The plateau-like behavior for $d \lesssim 1$ is only present for the active region. It results from the observed protrusions and thin bridges of active material connecting larger regions. The morphological asymmetry between the active and passive regions is even more pronounced if considered with respect to \rmic. Again the passive regions lead to longer tails if compared with the active regions, indicating larger more compact domains. With respect to the strength of activity the tails of the distributions shrink with increasing \activity  and the maximal values increase and shift to smaller sizes. In Figure \ref{fig:struct2} C we measure the connectivity of the active and passive regions. Considering the largest connected domain in each time instance of the dynamic equilibrium state, relating its area to the domain size and averaging this quantity, again shows the asymmetry between the active and the passive region in the active emulsion. Quantities of $0.5$ mean that the whole region is connected and thus correspond to bicontinuous structures. This is trivially realized for the macroscopic phase separated state for $\activity = 2.5$. With increasing activity, reaching the arrested coarsening states, this value decreases, e.g., due to the formation of droplets. However, within the active region it remains dominantly bicontinuous with values $\gtrsim 0.45$, whereas in the passive region the decrease is more sudden, corresponding to the stronger increase of the number of passive droplets in Figure \ref{fig:dynEq} G.    

\section{Conclusion}

Considering an established coarse-grained model for bacterial suspensions and extending it to two-phase systems, we explored the emerging morphological asymmetry of the phase-separating fluid mixture. Under the influence of activity in one bulk phase the characteristic coarsening process of such systems can arrest and form a dynamic equilibrium state. This active emulsion is characterized by interface fluctuations and asymmetries in geometric and topological properties, essentially more passive droplets in the active region than active droplets in the passive region, more compact domains in the passive region than in the active region, and a stronger connectivity of the active region than the passive region. We have demonstrated that all these phenomena result form self-organized flows in the active region. Due to an inverse energy cascade in the active region flow structures emerge and form condensates. The size of these mesoscopic vortices is determined by the locally available space in the active region. Therefore the morphology of the active emulsion and the flow field in the emulsion are tightly coupled. This is conceptionally similar to inertial turbulence in two-phases systems \cite{PhysRevA.23.3224,Bertetal_PRL_2005,perlekar2017two}. Instead of a balance of inertial and interfacial-tension forces leading to the KH scale \cite{Hinze_AICHE_1955}, the dynamic emulsion is characterized by a balance of active and interfacial-tension forces, similar to \cite{PhysRevLett.115.188302}.
But also similarities of the flow field can be quantified. As for inertial turbulence the energy spectrum at large scales indicates characteristics of hyperuniformity of the vorticity field. However, there are also differences, which lead to asymmetries of the morphology. Larger available space in the active region allows for the formation of larger vortices. These flow structures have higher energies, which generate stronger interface fluctuations and might initiate break-up. Such break-up likely occurs at protrusions. For active protrusions the locally available space in the active region is small, therefore pinch-off is less likely. For passive protrusions, the surrounding active region might be large, thus leading to stronger fluctuations which promotes pinch-off. As a results break-up preferentially leads to the formation of passive droplets. As a consequence, the passive region is less connected, leading to the topological asymmetry. Furthermore, as passive domains essentially just dissipate the energy impact at the interface, large compact passive domains are less affected, which explains the geometric asymmetry.  

While the observed asymmetries are similar to those found in active emulsions formed by systems of active nematic liquid crystals and passive isotropic fluids \cite{adkins2022interface,zhao2024asymmetric}, the physical origin differs. The considered mechanism does not explicitly require liquid crystalline order but only self-organization of flow. It can thus be considered as more general. While computationally explored for the case of an active/passive system, the proposed mechanism can be expected to also hold for heterogeneous active systems with long-ranged hydrodynamic flows, e.g., two-phase suspensions with internal boundaries separating different types of bacteria \cite{ajesh2022bacteria} or heterogeneous mixtures of cells with internal boundaries separating different cell types 
\cite{Gibbsetal_Science_2008,Buddingetal_JB_2009,Pattesonetal_NC_2018}. If different types of bacteria or different cell types are associated with different strength of activity, the heterogeneity in activity provides a way to break the symmetry in the flow field and as a consequence the morphology. This asks for experimental validation in order to explore the proposed mechanism as a way to form functional soft materials with tunable microstructure. \\

\noindent
{\bf Acknowledgements:} This work was supported by the German Research Foundation (DFG) through the project "Analysing structure-property relations in equilibrium and non-equilibrium hyperuniform systems" (project number VO 899/32-1). We further acknowledge computing resources at JSC under grant "MORPH" and at ZIH under grant WIR.


\bibliography{ActiveFlowLit}


\end{document}